\documentclass{jfm}
\usepackage{graphicx}
\usepackage{epstopdf, epsfig}
\usepackage{subcaption}
\usepackage[compatibility=false]{caption}
\usepackage{marvosym}
\usepackage{tikz}
\usepackage{listings} 
\PassOptionsToPackage{rgb}{xcolor}
\usepackage{pgfplots}
\usepgfplotslibrary{polar}
\pgfplotsset{compat=1.13}
\usepackage{epsf}
\usepackage{adjustbox}
\usepackage{float}
\usepackage[section]{placeins}
\usepackage{times}
\usepackage{amsmath}
\usepackage{amssymb}
\usepackage{bm}
\usepackage{ mathrsfs }
\usepackage{lscape}
\usetikzlibrary{external}
\tikzsetexternalprefix{./}
\newlength{\colourbarheight}
\renewcommand{\vec}[1]{\mathbf{#1}}

\shorttitle{Compressible Invariant Solutions in Cavity Flows}
\shortauthor{J. J. Otero, A. S. Sharma and R. D. Sandberg}

\title{Compressible Invariant Solutions In Open Cavity Flows}

\author{J. Javier Otero\aff{1}
  \corresp{\email{josejavier.otero.perez@gmail.com}},
  Ati S. Sharma\aff{1}
 \and Richard D. Sandberg\aff{2}}

\affiliation{\aff{1}Faculty of Engineering and the Environment, University of Southampton, Southampton SO17 1BJ, U.K.
\aff{2}Department of Mechanical Engineering, University of Melbourne, 3010 Victoria, Australia}

\begin{document}

\maketitle

\begin{abstract}

A family of compressible exact periodic solutions is reported for the first time in an open cavity flow setup. These are found using a novel framework which permits the computation of such solutions in an arbitrary complex geometry. 
The periodic orbits arise from a synchronised concatenation of convective and acoustic events which strongly depend on the Mach number. This flow-acoustic interaction furnishes the periodic solutions with a remarkable stability and it is found to completely dominate the system's dynamics and the sound directivity.
The periodic orbits, which could be called `exact Rossiter modes', collapse with a family of equilibrium solutions at a subcritical Hopf bifurcation, occurring in the quasi-incompressible regime.
This shows compressibility has a destabilising effect in cavity flows, which we analyse in detail.
By establishing a connection with previous 2D and 3D stability studies of cavity flows, we are able to isolate the effect of purely compressible two-dimensional flow phenomena across Mach number.
A linear stability analysis of the equilibria provides insight into the compressible flow mechanisms responsible for the instability.
A close look at the adjoint modes suggests that an eigenvalue merge occurs at a Mach number between 0.35 and 0.4, which boosts the receptivity of the leading mode and determines the onset of the unstable character of the system.
The effect of the choice of base flow over the transition dynamics is also discussed, where in the present case, the frequencies associated to the leading eigenmodes show a strong connection with the frequencies of the periodic orbits at the same Mach numbers.
\end{abstract}

\begin{keywords}
Nonlinear instability, bifurcation, flow stability, compressible flow, global modes.
\end{keywords}

\section{Introduction}
\label{sec:intro}

Flow states that remain exactly the same under some transformation can enrich our understanding of the basic physical mechanisms behind the flow phenomena we observe. Such solutions are known variously as invariant solutions, recurrent flows or exact coherent structures, and include equilibria, periodic orbits, and convecting versions of the same.
The calculation of such solutions promises an understanding of complex flow phenomena without the use of modelling or approximation. Where such solutions are unstable, chaotic flows can be thought of as a series of transitions from the proximity of one solution to another. This line of thinking has led to the view that statistical properties of these chaotic systems may be approximated using an expansion over (relative) periodic orbits \citep[e.g.][]{kawahara2001periodic,cvitanovic2005chaos}. Examples of previous investigations searching for steady states in incompressible flows used methods such as continuation \citep{continuation_keller}, selective frequency damping \citep{aakervik2006steady}, and an adjoint-based approach \citep{farazmand}. The numerical computation of exact steady solutions was extended to compressible flows by \cite{yamouni2013}, who used a Newton method. On the other hand, this type of work was first carried out with the aim of seeking non-steady states by \cite{nagata1990three}, and since then, the search of such flow solutions remained exclusively applied to incompressible flows. For further details of similar investigations, the reader is referred to  the reviews from \cite{doi:10.1146/annurev-fluid-120710-101228} and \cite{cvitanovic2013recurrent}. More recently, \cite{farazmand} extended the available methods to obtain steady and travelling wave solutions using an adjoint-based approach. With his framework, convergence to an exact solution is guaranteed regardless from the flow state used as initial condition. This is an advantage over numerical Newton-Raphson methods, which require a starting point sufficiently close to the sought solution for convergence, although the latter converges faster in such a case.

\subsection{Cavity Flows}
\label{subsec:cavity_flows}
We apply our newly developed framework to the well studied case of flow over a two-dimensional open cavity at a Reynolds number $Re_D=2000$, which is defined as $Re_D=\rho_\infty U_\infty D/ \mu_\infty$. The subscript $\infty$ indicates free-stream values and D represents the cavity depth. Despite the relatively simple geometry of this case, the limited computational resources available to early investigators forced them to focus on the incompressible flow mechanisms, and insights into the compressible events present in cavity flows were mostly accessible through experimental research. In one of these investigations, \cite{rossiter_modes} documented and studied in detail for the first time the self-sustained flow oscillations of compressible origin, which are now commonly known as Rossiter modes. The origin of these periodic events resides in Kelvin-Helmholtz instabilities which grow along the separated shear layer, impinging on the cavity's trailing edge. This flow impingement radiates an acoustic wave which also propagates upstream and, due to the high receptivity of the leading edge (see section \ref{subsec:stability_of_equilibria}), fuels the appearance of new shear layer instabilities. This Mach-number dependent flow-acoustic interaction triggers these new instabilities and governs the overall sound directivity.

One of the first relevant computational studies which simulated two-dimensional cavity flows using compressible DNS was carried out by \cite{rowley2002self}. They performed a large parametric study (changing $\Rey_D$, $M$, $L/D$, etc.) and documented in thorough detail the dynamics in each case. For low aspect ratio cavities (such as the present case) they found that the dynamics were governed by a shear layer (Rossiter) mode. At higher aspect ratios, the cavity flow abandons the shear layer mode and undergoes a transition towards a wake mode type of motion. In a parametric sense, our results extend their database further for cavities of aspect ratio $L/D=3$.

The stability of several open cavity flow configurations has been widely studied in past investigations. \cite{vassilios_gls} highlights this well-known flow geometry in his review of global linear stability. \cite{brs2008} carried out BiGlobal instability analysis \citep{theofilis2003algorithm} on various compressible 2D and 3D cavity flows, where for the three-dimensional cases they used the two-dimensional mean flow as the base state. There they discovered that the spanwise instabilities were independent of the Mach number, where their origin was of convective (rather than acoustic) nature. This finding motivated the three-dimensional stability analysis in cavity flows assuming incompressible flow, such as \cite{devicente2014}. From the neutral stability curves shown in \cite{meseguergarrido2014} (their figure 7), for an aspect ratio of $L/D=3$, the critical Reynolds number above which the three-dimensional cavity flow becomes unstable is $Re_D\approx800$. Hence, this implies that if all the flow configurations studied in this article were extended to a three-dimensional space, they would exhibit an unstable spanwise mode. In that case, the 2D and 3D modes would interact causing a frequency modulation respect to their isolated behaviour \citep{brs2008}.

Previous studies have tried to understand the behaviour of Rossiter modes in compressible cavity flows with a relatively high Reynolds number \citep[e.g.][]{kegerise2004mode}. Under such conditions, the two-dimensional convective instabilities dominate and exert a strong modulation over the Rossiter modes, which makes the understanding of these shear layer modes extremely difficult. For example, in the work of \cite{yamouni2013}, the fact that their cavity flow became more unstable with an increasing Mach number was reported as surprising. In that regard, we restrict our analysis to a two-dimensional space, where the Reynolds number is below the threshold where convective instabilities start to appear. With this choice, we may examine physical mechanisms governing the two-dimensional shear layer modes, which are exclusively compressible in origin.


\cite{kervik2007} investigated the linear stability of an incompressible steady solution of a two-dimensional open cavity flow. Despite their assumption of incompressible flow with a relatively low Reynolds number ($Re=350$), this equilibrium solution was unstable due to the cavity's large aspect ratio ($L/D\approx 25$), which caused the separated flow to undergo transition towards a wake mode type periodic cycle \citep{rowley2002self}. A similar study was carried out by \cite{luchini_2007}, this time investigating the instability in a 2D cylinder's wake, also using several incompressible steady solutions as the base flow for the stability problem. So far, the use of linear stability analysis using compressible invariant solutions has only been considered by \cite{yamouni2013}. With an open cavity flow with an aspect ratio of $L/D=1$ and the Reynolds number being $Re_D=7500$, their flow was already unstable in the incompressible regime, which prevented them from isolating the origin of the instability as a function of the Mach number.

In this article, we apply our framework to a two-dimensional open cavity flow at $Re_D = 2000$ and $M=0.5$ as a central case. As we will see, this particular flow solution naturally decays very slowly towards a limit cycle. Hence, we use our method to drive the flow state directly to the periodic orbit, skipping this long transient. After this reference periodic solution is computed, we use it as the initial guess to compute the equivalent periodic solution at neighbouring Mach numbers, keeping the Reynolds number $Re_D$ constant. The steady invariant solutions associated to some of these periodic orbits are also computed. The numerical method related to the computation of compressible invariant solutions is introduced in section \ref{sec:num_method}. Both families of equilibrium and periodic solutions are thoroughly detailed in sections \ref{sec:periodic_family} and \ref{sec:equilibria}, highlighting the effects of a changing Mach number over the flow dynamics. Moreover, a linear stability analysis is performed on these steady solutions in section \ref{sec:equilibria} in order to unveil the physical mechanisms which make these flow equilibria unstable for high Mach numbers.  
A dataset including the calculated equilibria and periodic solutions has been made available \citep{cavity_sol_dataset} and animations of the flow solutions are available in the online supplement.

\section{Numerical Method}
\label{sec:num_method}

The process of finding exact flow solutions is reduced to a simple optimisation exercise, where the optimisation parameters are the state variables $Q_0\left(\vec{x}\right)=Q\left(\vec{x},t_0\right)$ at the initialisation of the direct numerical simulation (DNS). The numerical framework used to obtain the exact flow solutions couples an in-house compressible DNS solver \citep[HiPSTAR -][]{rsand} with the L-BFGS optimisation algorithm \citep{lbfgsb}. Starting from an initial condition (i.e. an arbitrary flow snapshot), the DNS provides the value of a chosen cost function to be optimised, but also the gradients to all the control parameters. The value of the cost function alongside the gradients are fed to the optimisation algorithm, which returns a better estimate of initial state variables. This simple algorithm is then repeated until some stopping criterion is met, where the nature of the solution found by the algorithm (steady or time-periodic) depends exclusively on the chosen cost function.

The cavity flow simulation is arranged into a four-block setup, where the grid is of Cartesian type and continuous up to fourth order across blocks. With the coordinate origin at the lower-left corner of the cavity and using the cavity depth $D$ as the reference length, the domain ranges from -20 to 20 in the streamwise direction, and from 0 to 10 in the vertical direction; restricting the DNS resolution to the vicinity of the cavity. This discretisation leads to a total of 480 000 grid points, where the resolution of the cavity is $800 \times 300$ points. Note that this grid resolution is much higher than the one used by \cite{rowley2002self} and it has been shown to produce grid independent results in \cite{mythesis}. For a complete description of the flow governing equations the reader is referred to appendix \ref{app:gov_eq}.

\subsection{Steady Solutions}
\label{subsec:steady_solutions}

To find a steady flow solution, we require a cost function which penalises the change of the flow-field throughout the time integration respect to the initial state. Such function can be written as \begin{equation}
 \mathcal{J}\left(Q\left(\vec{x},t\right),Q_0\left(\vec{x},t_0\right)\right) = \int_{T}\int_{\Omega}\frac{1}{2}\left|Q\left(\vec{x},t\right) - Q_0\left(\vec{x},t_0\right)\right|^2 \mathrm{d}\vec{x} \mathrm{d}t,
\label{eq:steady_cost_f}
\end{equation}
where $T$ is the duration of the time integration, $\Omega$ is the computational domain and $\left|\cdot\right|$ indicates an appropriate norm. In order to drive the cost function towards its minimum using the L-BFGS method, it is necessary to compute the gradients of the cost function to the initial condition, which are written as
\begin{equation}
\frac{D\mathcal{J}}{DQ_0} = \frac{\partial \mathcal{J}}{\partial Q} \frac{\mathrm{d}Q}{\mathrm{d}Q_0} + \frac{\partial \mathcal{J}}{\partial Q_0}.
\label{eq:grad_fwd_steady}
\end{equation}
In this particular case, the gradients can be computed straight from the DNS without any further cost as\begin{equation}
\left. \frac{D\mathcal{J}}{DQ_0}\right|_{\vec{x}} = \int_{T}- \left[Q\left(\vec{x},t\right) - Q_0\left(\vec{x},t_0\right)\right] \mathrm{d}t,
\end{equation} since the first term of the right hand side of (\ref{eq:grad_fwd_steady}) cancels out. From a preliminary study, it was found that longer horizons provide better gradients with more information about the leading instability of the initial state.  Hence, increasing the length of this horizon resulted in a flow-field matching the initial state for a longer time. Note that this would only be the case when the steady solutions are unstable. Finally, the stopping criterion used consists in the variations of the cost function $\mathcal{J}$ being of the order of the numerical precision used. All the equilibria presented  later on were computed using their corresponding periodic orbit at the same Mach number as the starting point.

\subsection{Periodic Solutions}
\label{subsec_6:periodic_solutions}

In order to search for a time-periodic flow solution, we use less constrained cost function,
\begin{equation}
 \mathcal{J}\left(Q\left(\vec{x},T\right),Q_0\left(\vec{x},t_0\right)\right) = \int_{\Omega}\frac{1}{2}\left|Q\left(\vec{x},T\right) - Q_0\left(\vec{x},t_0\right)\right|^2 \mathrm{d}\vec{x}.
\end{equation}
Note that in contrast to equation (\ref{eq:steady_cost_f}) there is no time integral, so the cost is now just the squared norm of the difference between the initial and final flow states. Following the same reasoning as before, the gradients also follow directly from the DNS,
\begin{equation}
 \left.\frac{D\mathcal{J}}{DQ_0}\right|_{\vec{x}} = - \left[Q\left(\vec{x},T\right) - Q_0\left(\vec{x},t_0\right)\right].
\end{equation}
Here, the time horizon $T$ is not fixed and so also has to be optimised. One possible way to find a good initial estimate of $T$ is seeking a globally periodic pattern in the flow-field with distributed monitor points across the entire flow domain. After performing a Fourier transform of each individual  signals, all monitor points should show an energy peak at a common frequency, allowing an initial guess of $T$. To optimise the horizon length, instead of computing a time gradient with respect to the cost function introducing a new optimisation variable, $T$ is adjusted at every new iterate. This is achieved by overrunning the horizon $T$ by a minimal amount, and finding which nearby time-step has the minimum cost. For this approach to work, the initial time horizon guess must be close to the period of the final flow orbit. The convergence criterion used for these periodic solutions is $\mathcal{J}/\left\rVert Q_0 \right\rVert< 10^{-12}$.

\section{Periodic Solution at M=0.5}
\label{sec:PS_M050}

Before delving into the analysis of the evolution of periodic solutions across Mach number, we first describe the periodic orbit found for $M=0.5$ and $Re_D=2000$. This solution was found first and will be referred to when comparing other orbits. 
The analysis carried out in this section breaks down the periodic trajectory into more fundamental intervals. To characterise the periodic orbits, we define the norm
\begin{equation}
    \left\lVert \alpha \right\rVert = \int_{\Omega} \alpha\left(\vec{x},t\right)^2 W_i\left(\vec{x}\right)\text{d}\vec{x},
\label{eq:periodic_norm}
\end{equation}
where $\alpha$ is a space-time dependent flow quantity, and $W_i\left(\vec{x}\right)$ is a spatial function which is zero outside the vicinity of the cavity, preventing spurious sensitivity to the boundary conditions. Here, the function $W_i\left(\vec{x}\right)$ is defined as
\begin{equation}
W_i\left(\vec{x}\right) = \begin{cases}
 1 & \quad \text{if} \ \vec{x} \in \left[\left(-1.5,0\right),\left(4.5,4\right)\right]\\
 0 & \quad \text{if} \ \vec{x} \notin \left[\left(-1.5,0\right),\left(4.5,4\right)\right]\\
\end{cases}.
\end{equation}

\begin{figure}
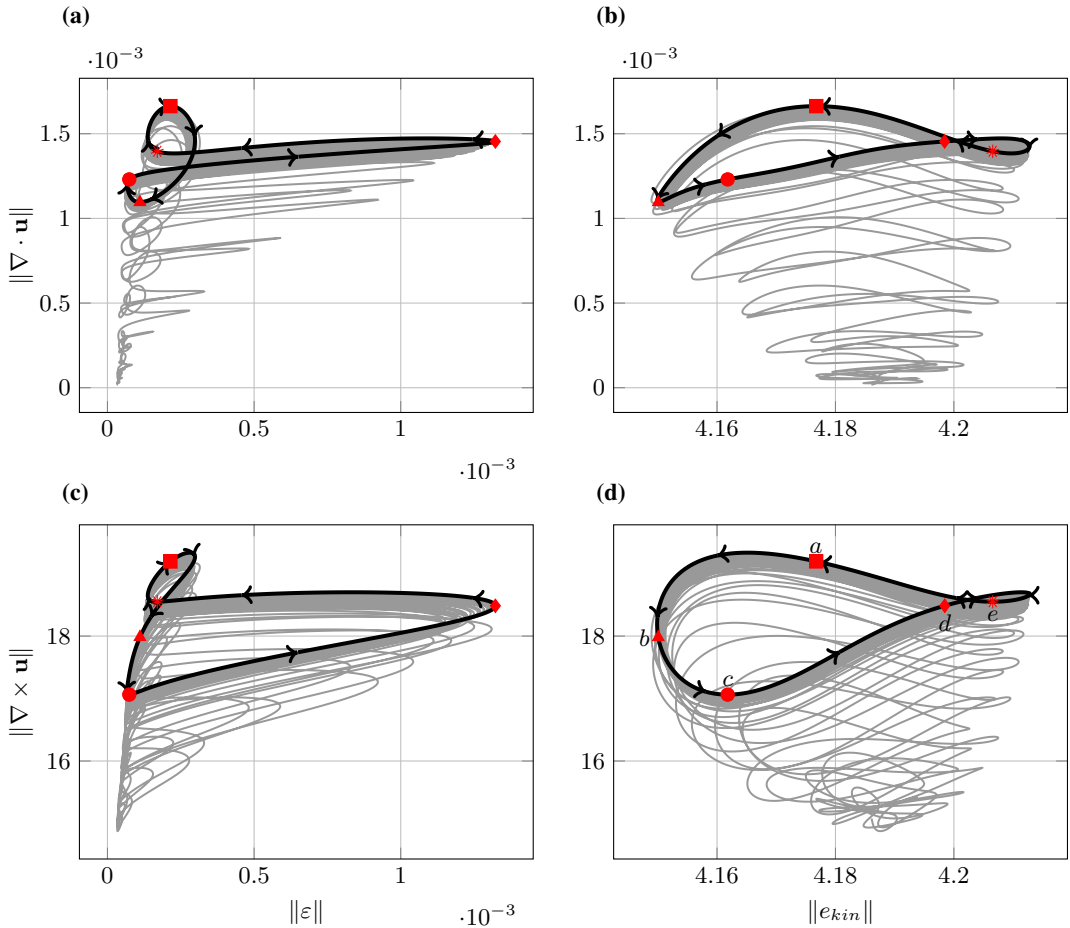

 \begin{subfigure}[b]{0.49\textwidth}
  \centering
  \subcaption{}
     \input{./Re_2000_M050_a}
     \label{fig:diss_vs_dil}
 \end{subfigure}
 ~\hspace{0.02\textwidth}
 \begin{subfigure}[b]{0.49\textwidth}
  \centering
  \subcaption{}
     \input{./Re_2000_M050_b}
     \label{fig:ke_vs_dil}
 \end{subfigure}\\
 \begin{subfigure}[b]{0.49\textwidth}
  \centering
  \subcaption{}
     \input{./Re_2000_M050_c}
     \label{fig:ke_vs_vort}
 \end{subfigure}
 ~\hspace{0.02\textwidth}
 \begin{subfigure}[b]{0.49\textwidth}
  \centering
  \subcaption{}
     \input{./Re_2000_M050_d}
     \label{fig:diss_vs_vort}
 \end{subfigure} 
 \caption{4D representation of the periodic orbit at $M=0.5$. The grey line shows the natural flow evolution from the vicinity of the initial condition towards the periodic orbit. The red symbols indicate key reference points (colour online).}
\label{fig:phase_portrait_Re_2000_M050}
\end{figure}

To analyse the physics of the periodic orbits, we have selected dilatation ($\nabla \cdot \vec{u}$), which highlights the compressible events, and viscous dissipation rate \citep[$\varepsilon$ - see][]{kundu_book}, which should emphasise the flow phenomena with strong shear, such as vortex merging. Additionally, we use vorticity ($\nabla \times \vec{u}$) and kinetic energy ($e_{kin}$), which are often used to characterise periodic orbits in incompressible flows. Figure \ref{fig:phase_portrait_Re_2000_M050} shows a phase portrait of the periodic solution for $Re_D=2000$ and $M=0.5$, projected onto these four variables. The locations of the key physical events occurring in the solution are highlighted with symbols. These symbols are labelled chronologically from $a$ to $e$ in figure \ref{fig:diss_vs_vort}, where $a$ indicates the vortex impingement on the trailing edge of the cavity, which is the instant where compressible effects reach their maximum. Also in figure \ref{fig:phase_portrait_Re_2000_M050}, the grey lines show how the flow evolves from the flow initial condition towards the periodic trajectory. It can be seen that the periodic orbit is an attractor which almost represents the complete behaviour of the flow. This follows as a consequence of the relatively low Reynolds number alongside a strong acoustic shear layer feedback mechanism which stabilises the flow onto this periodic behaviour. Hence, the flow lacks the sufficient energy to `jump' to another state and sits indefinitely in the close vicinity of this stable orbit.

\begin{figure}
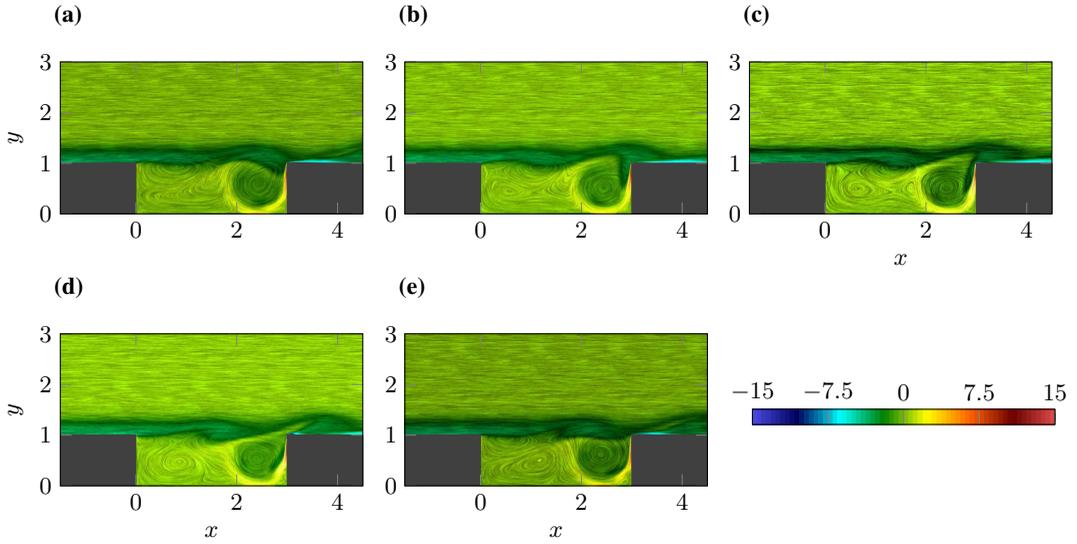

 \begin{subfigure}[b]{0.32\textwidth}
  \centering
  \subcaption{}
     \input{./Re_2000_M050_vort_a}
     \label{fig:vort_a}
 \end{subfigure}
 ~\hspace{0.00\textwidth}
 \begin{subfigure}[b]{0.32\textwidth}
  \centering
  \subcaption{}
     \input{./Re_2000_M050_vort_b}
     \label{fig:vort_b}
 \end{subfigure}
 ~\hspace{0.00\textwidth}
 \begin{subfigure}[b]{0.32\textwidth}
  \centering
  \subcaption{}
     \input{./Re_2000_M050_vort_c}
     \label{fig:vort_c}
 \end{subfigure}\\
 ~\hspace{0.00\textwidth}
 \begin{subfigure}[b]{0.32\textwidth}
  \centering
  \subcaption{}
     \input{./Re_2000_M050_vort_d}
     \label{fig:vort_d}
 \end{subfigure}
 ~\hspace{0.00\textwidth}
 \begin{subfigure}[b]{0.32\textwidth}
  \centering
  \subcaption{}
     \input{./Re_2000_M050_vort_e}
     \label{fig:vort_e}
 \end{subfigure} 
 ~\hspace{0.00\textwidth}
 \begin{subfigure}{0.32\textwidth}
 	\vspace{-2.1cm}
 	\centering
  	\input{./Re_2000_M050_vort_colourbar}
 \end{subfigure}
 \caption{Contours of instantaneous $z$-vorticity at the time steps represented by the red symbols in figure \ref{fig:phase_portrait_Re_2000_M050} (colour online). That is, \textbf{(a)} corresponds to the mark labelled with $a$, \textbf{(b)} with the mark labelled with $b$ and so on.}
\label{fig:vort_series}
\end{figure}

\begin{figure}
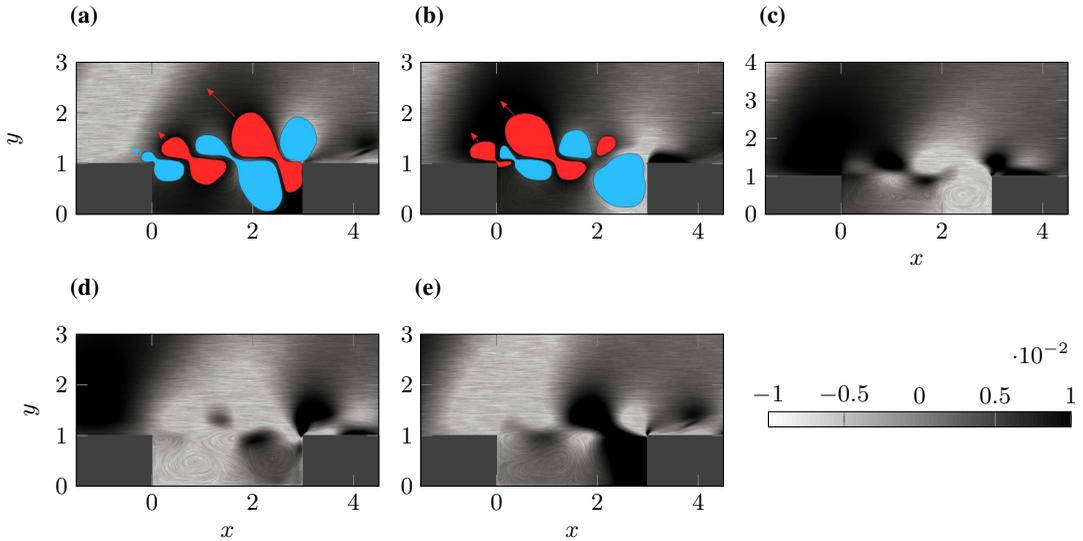

	\begin{subfigure}[b]{0.32\textwidth}
		\centering
		\subcaption{}
     	\input{./Re_2000_M050_dil_a_cartoon}
     	\label{fig:cartoon_a}
    \end{subfigure}
	~\hspace{0.00\textwidth}
 	\begin{subfigure}[b]{0.32\textwidth}
   		\centering
   		\subcaption{}
      	\input{./Re_2000_M050_dil_b_cartoon}
      	\label{fig:cartoon_b}
    \end{subfigure}     
 	~\hspace{0.00\textwidth}
 	\begin{subfigure}[b]{0.32\textwidth}
   		\centering
   		\subcaption{}
      	\input{./Re_2000_M050_dil_c}
      	\label{fig:dil_c}
    \end{subfigure}\\
	\begin{subfigure}[b]{0.32\textwidth}
		\centering
		\subcaption{}
     	\input{./Re_2000_M050_dil_d}
     	\label{fig:dil_d}
     \end{subfigure}
 	~\hspace{0.00\textwidth}
 	\begin{subfigure}[b]{0.32\textwidth}
   		\centering
   		\subcaption{}
      	\input{./Re_2000_M050_dil_e}
      	\label{fig:dil_e}
    \end{subfigure} 
 	~\hspace{0.00\textwidth}
 	\begin{subfigure}{0.32\textwidth}
 		\vspace{-2.5cm}
   		\centering
      	\input{./Re_2000_M050_dil_colourbar}
    \end{subfigure} 
    \caption{\textbf{(a)} and \textbf{(b)} - Illustration of the leading edge and shear layer dipole interaction at instants $a$ and $b$. \textbf{(c)} to \textbf{(e)} - Snapshots of the dilatation field at instants $c$, $d$ and $e$.} 
   \label{fig:dil_d_e}
\end{figure}

To further illustrate the flow behaviour, snapshots of the vorticity field are shown in figure \ref{fig:vort_series}, where the subfigures are ordered chronologically with the sub-caption matching the instants labelled in figure \ref{fig:diss_vs_vort}. At the instant $a$, the vortex located in the downstream end of the cavity impinges onto the trailing edge (figure \ref{fig:vort_a}). This vortex was also observed in, for example, \cite{brs2008}, and remains in that location throughout the entire periodic orbit. For this reason, it will be referred to as the stationary vortex. At this instant, the vortex is slightly stretched in the vertical direction, which is why it impinges on the trailing edge of the cavity. This stretching is the result of the previous shear layer vortex merging with this stationary vortex, which has lead to a low density (and high momentum) area on top of this stationary vortex. Hence, this vortex impingement radiates a low density acoustic wave. Additionally, another vortex is attached to the leading edge of the cavity. As mentioned, this is the instant where the compressible effects are higher, mostly, but not entirely, due to the vortex impingement occurring at the trailing edge of the cavity. Figure \ref{fig:cartoon_a} shows how the sound radiation is mainly generated in the upstream direction. The physical phenomenon is due to two aligned dipoles in perfect synchronisation. The strongest dipole is located at the trailing edge of the cavity and its origin is the impingement of the stationary vortex on the cavity's trailing edge. Whereas, the weaker dipole is caused by the clockwise rotation of the vortex currently attached to the leading edge of the cavity. Note that this dipole interaction partially cancels the sound radiation in the downstream direction. 

\begin{itemize}

\item[From $a$ to $b$] the intensity of the impingement decreases rapidly, which causes the trailing edge dipole to dissipate while the shear layer dipole gains in intensity. Additionally, the upstream vortex is detached from the leading edge of the cavity. As this vortex moves downstream, a new vortex is formed in the upstream lower corner of the cavity. This vortex rotates counter-clockwise and induces a downwash along the upstream vertical wall of the cavity, which expands the flow as it interacts with the incoming boundary layer. This results into a new dipole attached to the leading edge, which also contributes to the amplification of the acoustic wave reflected from the trailing edge at $a$ (figure \ref{fig:cartoon_b}). When the flow reaches $b$, dilatation and kinetic are at their minimum values (figure \ref{fig:ke_vs_dil}). At this point, the trailing edge dipole has vanished, leaving temporarily as leading noise sources the shear layer and leading edge dipoles.

\item[From $b$ to $c$] the vortex in the shear layer keeps moving forward, with its corresponding dipole now decaying in intensity, but still further amplifying the upstream propagating sound wave. The main contributor to this amplification is now the dipole at the leading edge. The instant $b$ can be seen as the start of a compressible interaction between the shear layer and stationary vortices, where their counter-rotating behaviour compresses the flow between the two vortices which creates a high density spot. Note that the norm of the dilatation field increases due to this phenomenon (figure \ref{fig:diss_vs_dil}). Also, the norm of vorticity (figure \ref{fig:ke_vs_vort}) reaches minimum values due to the low interaction of vortical structures with the trailing edge (figure \ref{fig:vort_c}).

\item[From $c$ to $d$] the shear layer and stationary vortices collide and start a merging process. This interaction gains in intensity continuously until the orbit arrives at $d$. This phenomenon is reflected in figure \ref{fig:diss_vs_dil}, where the viscous dissipation rate increases suddenly and reaches a maximum at the instant $d$. The flow compression occurring downstream from the shear layer vortex has grown further and impinges on the trailing edge at $d$. This impingement reflects a high-density wave that also propagates upstream. At the leading edge of the cavity, a new vortex starts forming at $c$ due to the suction caused by the vortex at the upstream lower corner of the cavity. Again, the counter-rotation between the leading edge vortex and the shear layer vortex create another high-density spot in between them, which further amplifies the upstream propagating high-density wave. The instantaneous dilatation field shows again two dipoles in the shear layer and trailing edge in perfect synchronisation radiating noise in the upstream direction (figure \ref{fig:dil_d}). Note that this time, the dipoles show opposite sign with respect to $a$. Hence, $d$ could be seen as the phase counterpart of $a$.

\item[From $d$ to $e$,] soon after $d$, the core of the shear layer vortex gets absorbed by the stationary vortex. As the vortex merging completes, the norm of the viscous dissipation rate experiences a sudden drop (figure \ref{fig:diss_vs_dil}). Meanwhile, the leading edge vortex continues growing in size. As a result of this merging, the stationary vortex is stretched in the vertical direction as it keeps rotating clockwise towards the trailing edge, which causes the trailing edge dipole to slowly invert in sign. As well, the shear layer dipole which keeps contributing to the upstream sound radiation. Also, the clockwise rotation of the leading edge vortex pushes the flow upwards along the upstream vertical cavity wall compressing the flow. This phenomenon is observed in the dilatation field (figure \ref{fig:dil_e}) as a weak dipole located at the leading edge, which keeps amplifying the upstream travelling wave. Note that this same amplification mechanism occurred with opposite phase in $c$ (figure \ref{fig:cartoon_b}). 

\item[From $e$ to $a$,] the stationary vortex starts interacting with the trailing edge as it keeps rotating clockwise. This interaction is also observed in figure \ref{fig:diss_vs_dil}, where the norm of the dilatation field increases monotonically from $e$ to $a$. During this interval, the leading edge vortex has grown in size considerably, up to about half cavity length, just before it detaches from the leading edge again in $a$. These two physical phenomena occur in perfect synchronisation, leading to the double dipole, which radiates the acoustic wave upstream.

\end{itemize}

\section{Family of Periodic Solutions Across Mach Number}
\label{sec:periodic_family}

\begin{figure}
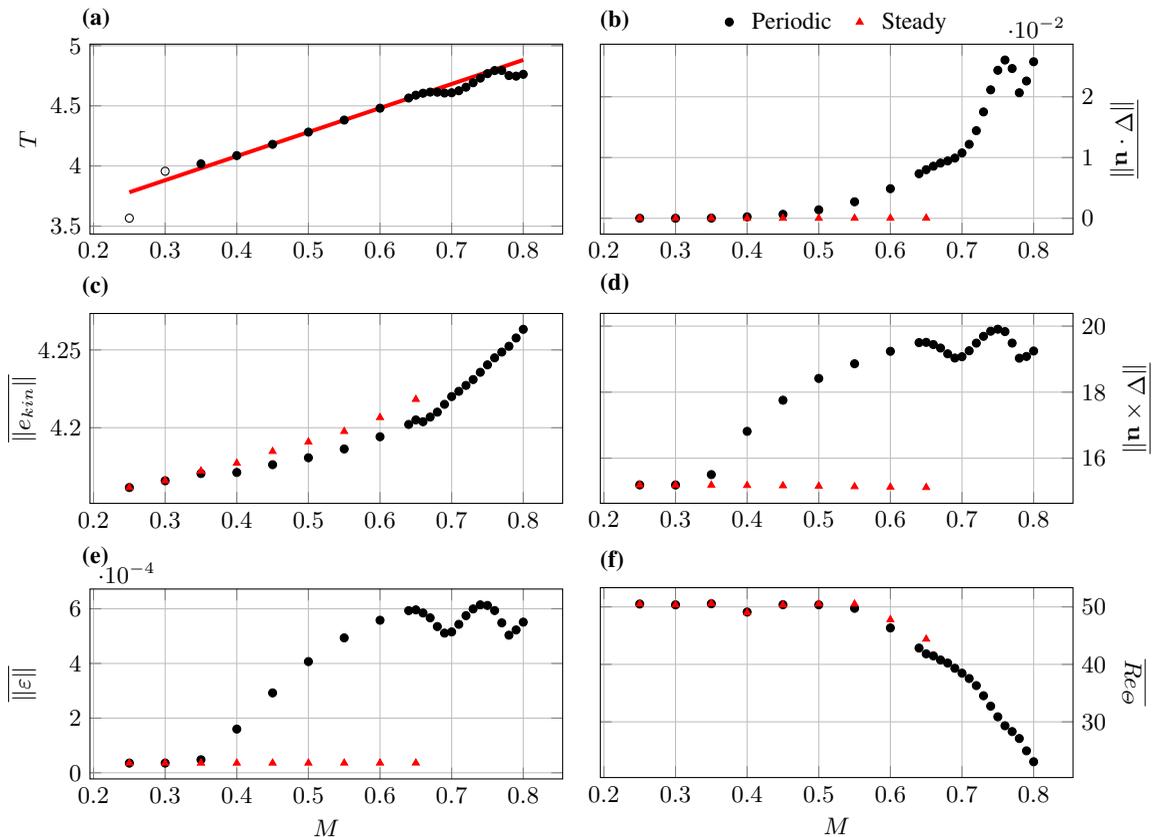

 \begin{subfigure}{0.49\textwidth}
  	\centering
 	\subcaption{}
 	\vspace{-0.25cm}
    \input{./T_vs_M}
    \label{fig:T_vs_M}
 \end{subfigure}
 ~\hspace{0.0\textwidth}
 \begin{subfigure}{0.49\textwidth}
 	\vspace{0.06cm}
  	\centering
  	\subcaption{}
  	\vspace{-0.65cm}
    \input{./dil_vs_M}
    \label{fig:dil_vs_M}
 \end{subfigure}\\
 \begin{subfigure}{0.49\textwidth}
  	\centering
  	\subcaption{}
  	\vspace{-0.05cm}
    \input{./ke_vs_M}
    \label{fig:ke_vs_M}
 \end{subfigure}
 ~\hspace{0.0\textwidth}
 \begin{subfigure}{0.49\textwidth}
  	\centering
  	\subcaption{}
  	\vspace{-0.05cm}
    \input{./vort_vs_M}
    \label{fig:vort_vs_M}
 \end{subfigure}\\
 \begin{subfigure}{0.49\textwidth}
 	\vspace{0.03cm}
  	\centering
  	\subcaption{}
  	\vspace{-0.325cm}
    \input{./diss_vs_M}
    \label{fig:diss_vs_M}
 \end{subfigure}
  ~\hspace{0.0\textwidth}
 \begin{subfigure}{0.49\textwidth}
 	\vspace{0.1cm}
  	\centering
  	\subcaption{}
  	\vspace{-0.05cm}
    \input{./Re_mom_vs_M}
    \label{fig:Re_thet_vs_M}
 \end{subfigure}
 \caption{\textbf{(a)} - Time period of the orbits as a function of Mach number. The red solid line shows the periods predicted by Rossiter's formula. \textbf{(b)} to \textbf{(f)} - Time-averaged quantities for the periodic orbits (black circles) and steady solutions (red triangles) across Mach number. The open circles represent purely numerical output of the algorithm without any physical interpretation. Colour online.}
 \label{fig:Mach_range}
\end{figure}

The non-dimensional character of our numerical framework permits us to use a periodic orbit with Mach number $M_0$ as an initial guess in the search of a new periodic orbit in neighbouring Mach numbers $M_0 \pm \delta$. In particular, the orbits were continued from the periodic trajectory found at $M=0.5$, both in ascending and descending order with increments of 0.05 in Mach number. Given the large variations in the flow quantities plotted in figure \ref{fig:Mach_range} for Mach numbers above 0.65, the step size in Mach number was reduced to 0.01. The range of Mach numbers studied covers $M=0.25$ to $M=0.8$. At the lower end ($M\gtrsim 0.35$), the periodic solution ceases to exist due to the low compressibility of the system and leads to a steady state. This phenomenon is also reflected in figure \ref{fig:Mach_range}, where the steady (see later section \ref{sec:equilibria}) and periodic solutions are seen to collapse at this lower Mach number regime.  On the other hand, for immediately higher Mach numbers, the interaction between the compressible and convective phenomena shown previously for the $M=0.5$ solution results in a family of periodic solutions which are stable.

\newlength\machfigheight
\setlength\machfigheight{2.5cm}

Figure \ref{fig:Mach_range} shows the period $T$ alongside other time-averaged quantities of interest as functions of Mach number. The periods shown in figure \ref{fig:T_vs_M} are compared with the predictions calculated using Rossiter's semi-empirical formula \citep{rossiter_modes}
\begin{equation}
T_M = \frac{L}{U_\infty} \frac{M+1/\kappa_M}{n-\gamma_M},
\label{eq:rossiter_formula}
\end{equation}
where $n$ is integer, $L$ is the cavity length and $\kappa_M$ and $\gamma_M$ are empirical constants\footnote{These empirical constants were calculated based on the periods obtained for the periodic orbits at Mach numbers 0.5 and 0.55. After solving the system of two equations with $\kappa_M$ and $\gamma_M$ as unknowns, we arrive to $\kappa_M = 0.6096$ and $\gamma_M = 0.5003$.}. These results correspond to the second cavity mode ($n=2$ - see subsection \ref{subsec:mom_t_and_s} for a description of the cavity mode selection mechanism). The agreement of the periods from the orbits presented in the present work and the predictions calculated using Rossiter's formula is remarkably good in the central section of figure \ref{fig:T_vs_M} ($0.35 \leq M \leq 0.65$). This particular range of Mach numbers shows a smooth and monotonic behaviour across all the plots shown in figure \ref{fig:Mach_range}. Note that the solutions at $M=0.25$ and $M=0.3$ are steady flow solutions, where the periods shown in \ref{fig:T_vs_M} are purely numerical outputs obtained by the algorithm without physical interpretation. As mentioned in the previous section for the $M=0.5$ orbit, this family of periodic solutions arise from the self-sustained compressible feedback mechanism characteristic of cavity flows. The vortex impingement on the trailing edge of the cavity radiates an upstream travelling acoustic wave that interacts with the oncoming new shear layer vortex. These two phenomena mutually benefit from each other due to their phase synchronisation along the entire orbit. When moving away from the reference $M=0.5$ solution in Mach number, the change in the propagating speed of the upstream travelling acoustic wave results into a phase modification of the acoustic wave shear layer interaction. For clarity in the following analysis, we refer to a synchronised interaction as `in phase', where the dipoles associated with the shear layer vortex and vortex impingement at the trailing edge have the same sign and enhance the upstream travelling acoustic wave (for example figures \ref{fig:cartoon_a} or \ref{fig:dil_d}). Contrarily, we say that the interaction occurs `out of phase' when these two dipoles have opposite sign and lessen the intensity of the radiated sound wave. The smooth and monotonic increase of the quantities shown in figure \ref{fig:Mach_range}, suggest that the higher flow compressibility associated with a higher Mach number dominates the average behaviour of the periodic orbits up to $M=0.65$. From this point onwards, the phase of the interaction appears to become a dominant phenomenon, leading to the oscillatory behaviour of the mean quantities observed mainly in figures \ref{fig:vort_vs_M} and \ref{fig:diss_vs_M}. Additionally, the rate of increase in average dilatation reduces from $M=0.65$ to $M=0.7$ due to the opposite phase of the acoustic wave and the shear layer dipoles, which partially cancels out the compressible phenomena. Note that this phase coupling might also vary as a function of Reynolds number (number of vortices in the shear layer) and cavity length (travelling distance of the upstream propagating acoustic wave). As the acoustic wave keeps decreasing its propagating speed (increasing the Mach number), the phase of the acoustic event becomes favourable again for the two physical mechanisms to work in synchronisation. This behaviour is reflected as a pronounced increase in the average dilatation from $M=0.7$ to $M=0.76$. As discussed later, the higher flow compressibility allows the shear layer not only to modulate the upstream travelling acoustic wave but also to radiate sound of comparable magnitude. This phenomenon, alongside its phase synchronisation with the already existing acoustic wave, is responsible for the sudden and steep changes in the periodic orbits above $M=0.75$ observed in figure \ref{fig:Mach_range}. Moreover, it is also worth tracking the evolution across Mach number of the boundary layer's momentum thickness $\Theta$ at the flow separation point, which is often used to characterise this type of flow. Variations in this particular quantity produce considerable changes in the amplitude of the above-described shear layer oscillations, and also slight modulations in the characteristic frequency. In particular, the amplitude of these oscillations appears to increase with rising $L/\Theta$, expressing its maximum differences in terms of $\overline{\left\lVert e_{kin} \right\rVert}$. Hence, the almost monotonic decrease in $Re_\Theta$ for $M>0.5$ shown in figure \ref{fig:Re_thet_vs_M} agrees with the steep increase in $\overline{\left\lVert e_{kin} \right\rVert}$ shown in figure \ref{fig:ke_vs_M} for the highest Mach number orbits. Furthermore, the trajectory's period also shows a close relation with $Re_\Theta$, where a decrease in $L/\Theta$ (increase in $Re_\Theta$) yields a longer period (see section \ref{subsec:mom_t_and_s} and also appendix \ref{appA}). Especially for $M>0.65$, an opposite oscillatory behaviour to the orbit's period as a function of Mach number (figure \ref{fig:T_vs_M}) is reflected in figure \ref{fig:Re_thet_vs_M}. More precisely, Mach number ranges which present a steep increase in the orbit's period correspond to a steep decrease in $Re_\Theta$ at that same Mach number range. Hence, it appears that the optimisation algorithm uses the incoming boundary layer thickness to balance the frequency of each flow trajectory, maintaining it within the proximity of the predictions from Rossiter's formula (\ref{eq:rossiter_formula}).

\begin{figure}
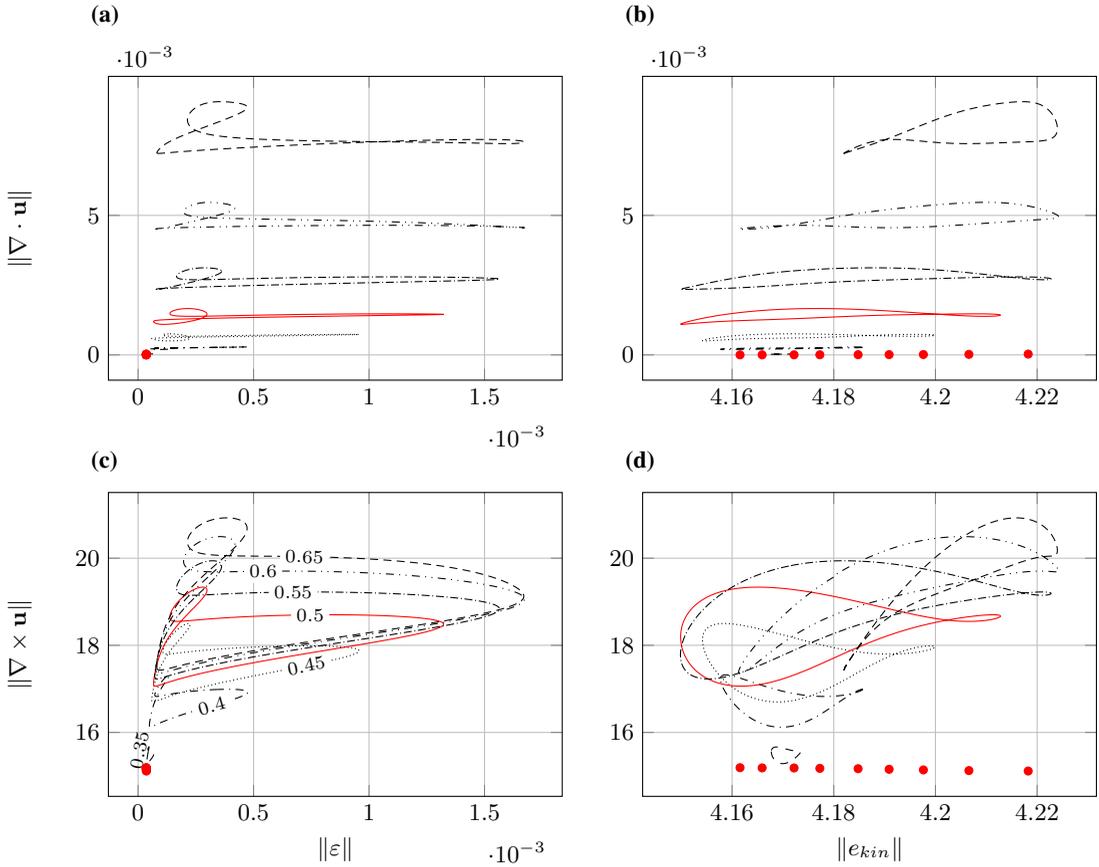

 \begin{subfigure}[b]{0.49\textwidth}
  	\centering
  	\subcaption{}
    \input{./diss_vs_dil_all}
    \label{fig:diss_vs_dil_all}
 \end{subfigure}
 ~\hspace{0.02\textwidth}
 \begin{subfigure}[b]{0.49\textwidth}
  	\centering
  	\subcaption{}
    \input{./ke_vs_dil_all}
    \label{fig:ke_vs_dil_all}
 \end{subfigure}\\
 \begin{subfigure}[b]{0.49\textwidth}
  	\centering
  	\subcaption{}
    \input{./diss_vs_vort_all}
    \label{fig:diss_vs_vort_all}
 \end{subfigure}
 ~\hspace{0.02\textwidth}
 \begin{subfigure}[b]{0.49\textwidth}
  	\centering
  	\subcaption{}
    \input{./ke_vs_vort_all}
    \label{fig:ke_vs_vort_all}
 \end{subfigure} 
 \caption{4D representation of the family of solutions across Mach number, ranging from 0.25 to 0.65. The equilibrium solutions from Mach numbers 0.25 to 0.65 (ordered with increasing kinetic energy) are represented with red dots.}
\label{fig:phase_portrait_Re_2000_all}
\end{figure}

\subsection{Low Mach number regime}

\begin{figure}
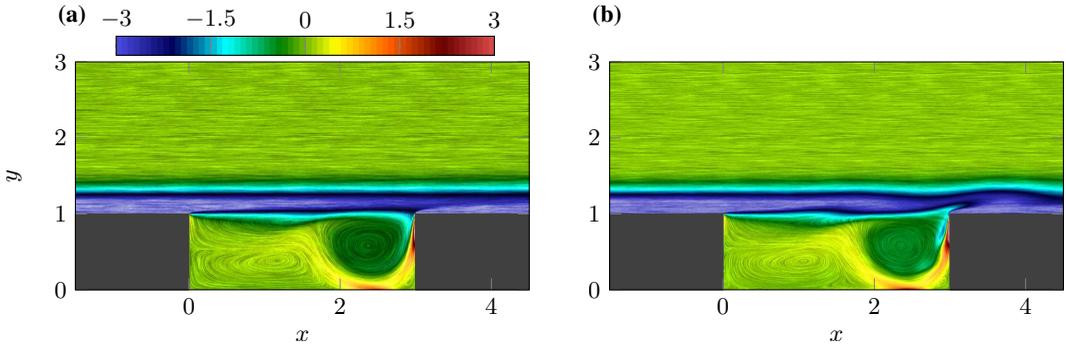

 \begin{subfigure}[b]{0.49\textwidth}
  	\centering
  	\subcaption{}
    \input{./Re_2000_M025_vort}
    \label{fig:M025_vort}
 \end{subfigure}
 ~\hspace{0.02\textwidth}
 \begin{subfigure}[b]{0.49\textwidth}
  	\centering
  	\subcaption{}
    \input{./Re_2000_M035_vort}
    \label{fig:M035_vort}
 \end{subfigure}\vspace{-4.60cm}
 \begin{subfigure}[b]{0.49\textwidth}
  	\centering
  	\input{./M_025_035_colourbar}
 \end{subfigure}\vspace{3.75cm}
 \caption{Instantaneous contours of $z$-vorticity at Mach numbers 0.30 \textbf{(a)} and 0.35 \textbf{(b)}.}
 \label{fig:M_025_035}
\end{figure}

Figure \ref{fig:phase_portrait_Re_2000_all} shows the phase portraits of the periodic orbits up to $M=0.65$. Note that three of the main features described in the previous section can be identified in figure \ref{fig:diss_vs_dil_all}. Instants $a$ and $b$ (vortex impingement, and shear layer vortex moving forward before interacting with the stationary vortex, respectively) correspond to the local maximum and minimum values of dilatation on the low viscous dissipation zone of the orbit. The $c$ in figure \ref{fig:phase_portrait_Re_2000_M050} (at the beginning of the merging of the shear layer and stationary vortices), merges with $b$ for Mach numbers above 0.5, as the difference in flow and sound velocity reduces. On the other hand, the instant $d$ (maximum intensity of the vortex merging) always follows the absolute maximum value of viscous dissipation rate, peaking at $M=0.6$. Interestingly, at Mach numbers 0.35 and 0.4 the strongest compressible event (maximum dilatation) is the vortex merging. After further increasing the Mach number, the maximum value of the norm of dilatation shifts to the instant corresponding with the vortex impinging on the trailing edge. This transition is strongly related to the physical mechanisms that cause the bifurcation of this family of periodic solutions from the steady solutions. 
As shown in figure \ref{fig:M025_vort}, the solution at $M=0.30$ sits at perfect equilibrium, where contrarily, a periodic trajectory exists for immediately higher Mach numbers (figure \ref{fig:phase_portrait_Re_2000_all}). Figure \ref{fig:M035_vort} shows the instantaneous vorticity field of the $M=0.35$ orbit at the point where the norm of the dilatation field is highest. When the steady solution becomes unstable, the weak leading edge vortex gets absorbed by the stationary vortex. During this merging process, similarly to the $M=0.5$ solution, the counter rotation of the two vortices originates a local flow compression, which is located downstream from the leading edge vortex (or also shear layer vortex). This phenomenon is the responsible for the highest norm of the dilatation field and coincides in time with the also highest norm of the viscous dissipation. From a density field perspective, this vortex counter-rotation leads to a high density spot sitting amongst the two vortices, whereas the vortices have an associated low density area on top of each one of them. Also note that the flow-field shown in figure \ref{fig:M035_vort} is the $M=0.35$ equivalent of the one shown in figure \ref{fig:vort_d} for the $M=0.5$ case. The reason why the vortex impingement is not as relevant from the compressible point of view is that the shear layer vortex is not strong enough to endure the orbit of the stationary vortex and it dissipates very rapidly. However, as the Mach number increases, the leading edge vortex gains in strength, which eventually leads to the low-density area impingement as the flow event with the highest compressibility. In addition, it is worth pointing out that the absolute minima of vorticity shown in figure \ref{fig:diss_vs_vort} represent the start of the merging amongst the shear layer and stationary vortices (instant $c$). Moreover, figure \ref{fig:ke_vs_vort_all} shows how the kinetic energy varies significantly with Mach number. Even though this is partially induced by the drop in $\Theta$, the overall change in the shape of the projection of the orbit also suggests that the phase of the interaction amongst the shear layer and acoustic wave begins to dominate the flow behaviour. Furthermore, the maximum value of the kinetic energy norm appears to reach saturation from $M=0.6$ to $M=0.65$ which shows that this interaction is shifting towards an out of phase synchronisation.

\begin{figure}
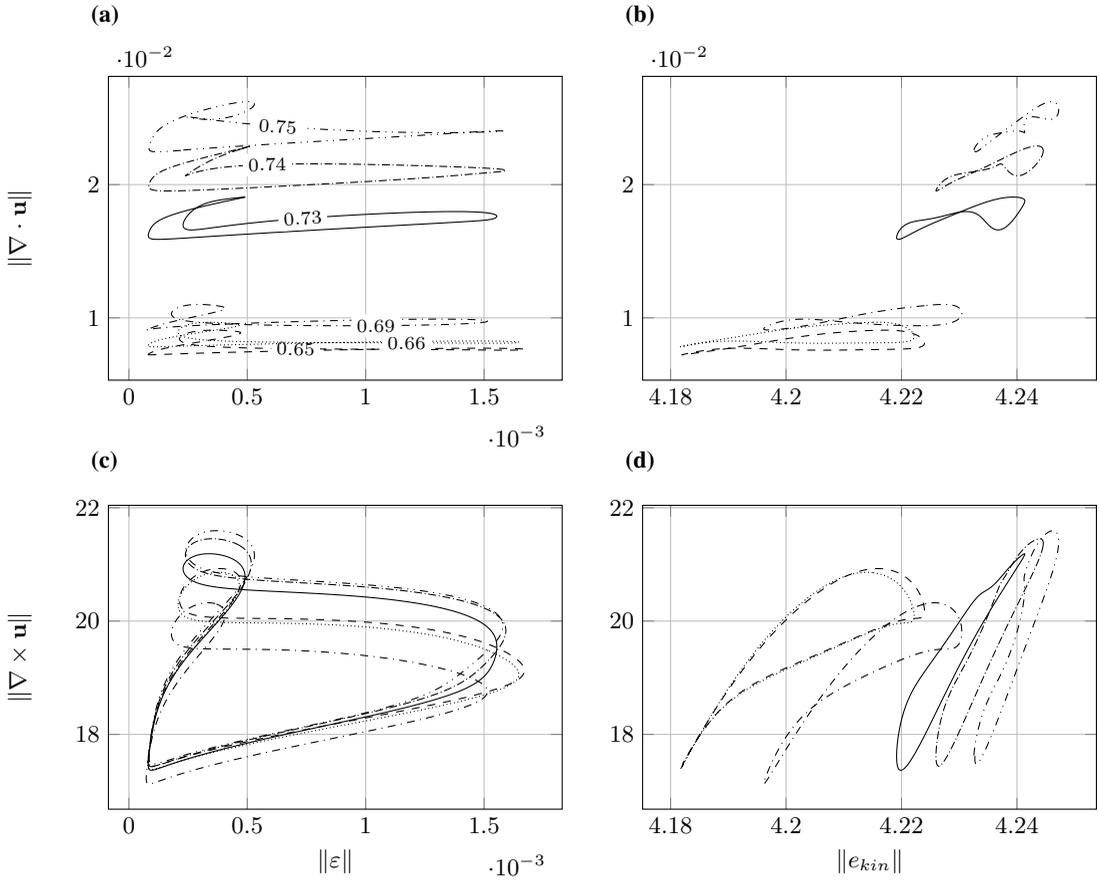

 \begin{subfigure}[b]{0.49\textwidth}
 	\centering
  	\subcaption{}
    \input{./diss_vs_dil_all2}
    \label{fig:diss_vs_dil_all2}
 \end{subfigure}
 ~\hspace{0.02\textwidth}
 \begin{subfigure}[b]{0.49\textwidth}
  	\centering
  	\subcaption{}
    \input{./ke_vs_dil_all2}
    \label{fig:ke_vs_dil_all2}
 \end{subfigure}\\
 \begin{subfigure}[b]{0.49\textwidth}
  	\centering
  	\subcaption{}
    \input{./diss_vs_vort_all2}
    \label{fig:diss_vs_vort_all2}
 \end{subfigure}
 ~\hspace{0.02\textwidth}
 \begin{subfigure}[b]{0.49\textwidth}
  	\centering
  	\subcaption{}
    \input{./ke_vs_vort_all2}
    \label{fig:ke_vs_vort_all2}
 \end{subfigure} 
 \caption{4D representation of the family of solutions across Mach number, ranging from 0.65 to 0.75.}
 \label{fig:phase_portrait_Re_2000_all2}
\end{figure}

\subsection{Phase-dominated Mach number regime}

\begin{figure}
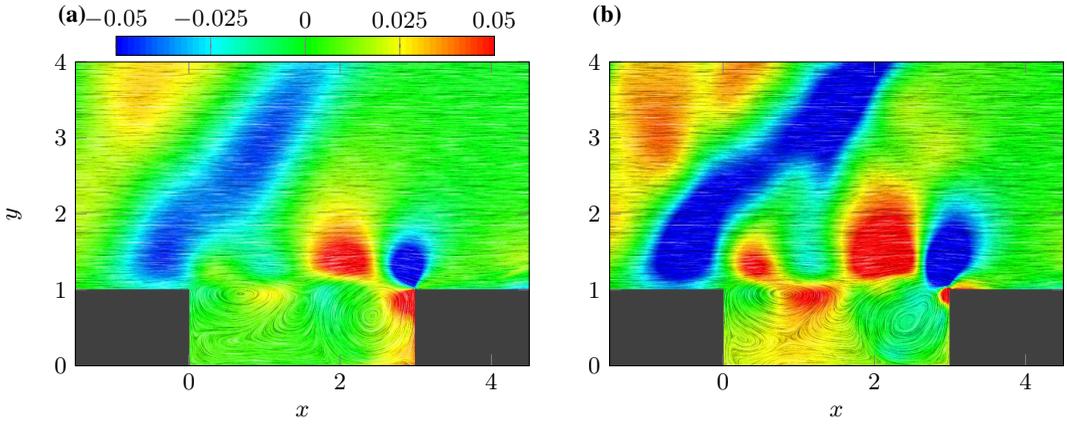

 \begin{subfigure}[b]{0.49\textwidth}
  \centering
  \subcaption{}
     \input{./Re_2000_M065_dil_max}
     \label{fig:dil_65_max}
 \end{subfigure}
 ~\hspace{0.02\textwidth}
 \begin{subfigure}[b]{0.49\textwidth}
  \centering
  \subcaption{}
     \input{./Re_2000_M074_dil_max}
     \label{fig:dil_74_max}
 \end{subfigure}\vspace{-5.6cm}
 \begin{subfigure}[b]{0.49\textwidth}
  \centering
  \input{./M_065_074_dil_max_colourbar}
 \end{subfigure}\vspace{5.5cm}
 \caption{Snapshots of the dilatation field at the maximum value of the norm of dilatation for Mach numbers 0.65 \textbf{(a)} and 0.74 \textbf{(b)}.}
\label{fig:dil_max_65_and_74}
\end{figure}

Figure \ref{fig:phase_portrait_Re_2000_all2} shows the phase-dominated Mach number range ($0.65<M<0.75$). Unlike the lower Mach number orbits represented in figure \ref{fig:phase_portrait_Re_2000_all}, the averaged norms of vorticity and viscous dissipation rate do not increase monotonically with Mach number (figure \ref{fig:Mach_range}). Instead, they describe an oscillatory behaviour related to the phase of the acoustic wave shear layer interaction. Figure \ref{fig:diss_vs_dil_all2} highlights how the phase of this interaction modifies the orbit at the vortex impingement on the trailing edge. As mentioned earlier in section \ref{sec:PS_M050}, this phenomenon is represented as the top left loop observed in the periodic orbits shown in figure \ref{fig:diss_vs_dil_all2}, which slowly unfolds as the Mach number increases. For the orbit at $M=0.74$, the beginning and the maximum intensity of the vortex impingement are distinctively represented as two sharp corners. Overall, the presence of more abrupt and complex features in these phase plots is strongly linked to higher flow compressibility, which permits the appearance of new dominant flow mechanisms present in the flow. To give insight into the interaction between the shear layer and the upstream travelling acoustic wave, figure \ref{fig:dil_max_65_and_74} shows the dilatation field at the maximum of the norm of dilatation (instant $a$) for Mach numbers 0.65 and 0.74. At figure \ref{fig:dil_65_max}, the interaction occurs in opposite phase, where the shear layer dipole induces a slight curvature in the upstream propagating sound wave. As the Mach number keeps increasing (figure \ref{fig:dil_74_max}) the speed of sound gets reduced compared to the flow velocity, bringing this interaction further out of phase. On the other hand, the greater flow compressibility in this scenario, permits the shear layer dipole to grow and it now radiates sound of comparable magnitude to the vortex impingement in the cavity's trailing edge. In essence, this shear layer dipole has progressed from slightly modifying the upstream propagating acoustic wave, to generating its own sound wave of similar magnitude. Hence, as the speed of sound decreases and the magnitude of the shear layer dipole grows, the combination of the two acoustic waves slowly shifts the main sound radiation towards a more vertical direction. For comparative purposes, these dilatation contours from these figures show a remarkable agreement with the ones shown in \cite{rowley2002self}, despite their differing configuration of $L/D=2$.

 Returning to figure \ref{fig:phase_portrait_Re_2000_all2}, the vortex impingement now sits as the highest value of dilatation and also in the vicinity of the highest kinetic energy, which differs from the lower Mach number range shown in figure \ref{fig:phase_portrait_Re_2000_all}. Similarly to the previous Mach number range, the maximum intensity of the vortex merging is also identified as the instant of maximum viscous dissipation rate in figures \ref{fig:diss_vs_dil_all2} and \ref{fig:diss_vs_vort_all2}. These two figures also show the shear layer vortex travelling downstream, right after detaching from the leading edge, as the minimum point in dilatation, vorticity and viscous dissipation. Shortly before that instant, the projected trajectory over the viscous dissipation and vorticity appears briefly to become independent from Mach number, where all the trajectories collapse. Also, as seen before in figure \ref{fig:ke_vs_M}, the average kinetic energy follows an increasing quasi-linear trend up to $M=0.8$. This is also reflected in figures \ref{fig:ke_vs_dil_all2} and \ref{fig:ke_vs_vort_all2}, where the horizontal displacement of the orbits as a function of Mach number is considerably higher than observed in figure \ref{fig:ke_vs_vort_all}. This phenomenon was found to be strongly related with the substantial drop in $\Theta$ at these periodic solutions.

\begin{figure}
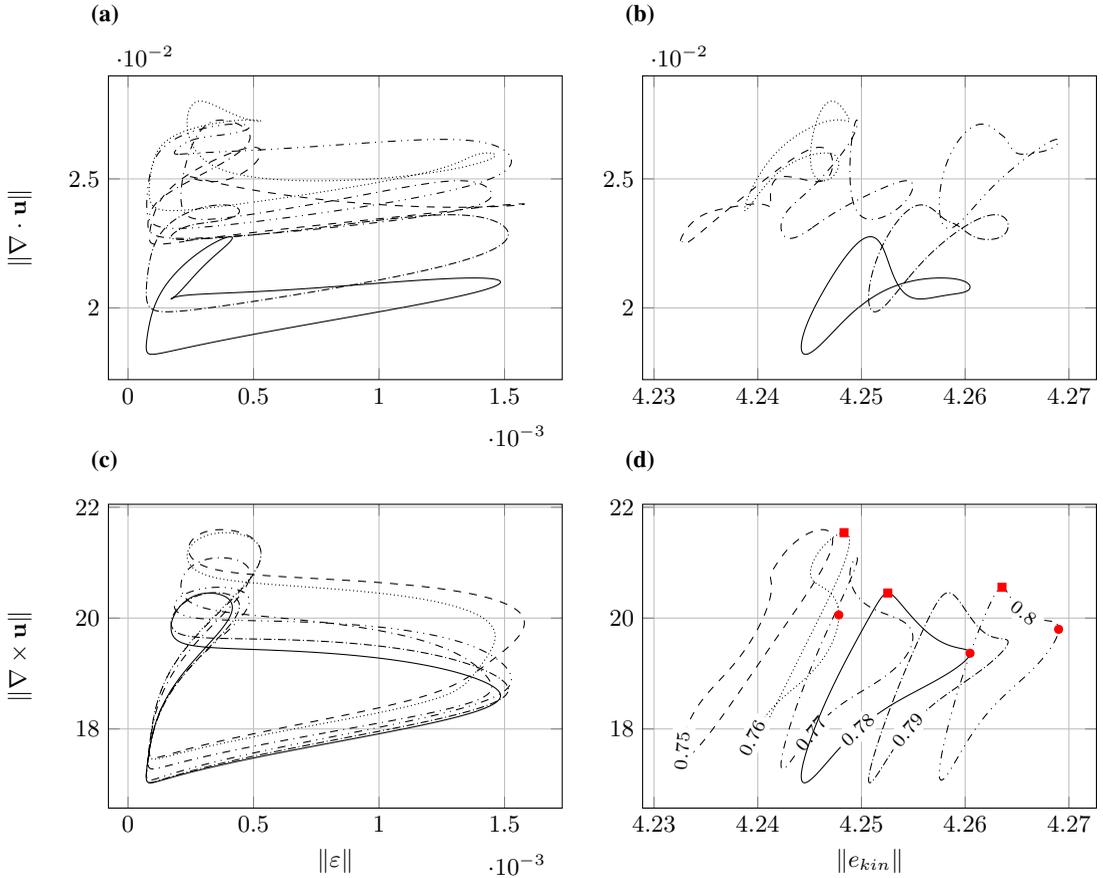

 \begin{subfigure}[b]{0.49\textwidth}
  \centering
  \subcaption{}
     \input{./diss_vs_dil_all3}
     \label{fig:diss_vs_dil_all3}
 \end{subfigure}
 ~\hspace{0.02\textwidth}
 \begin{subfigure}[b]{0.49\textwidth}
  \centering
  \subcaption{}
     \input{./ke_vs_dil_all3}
     \label{fig:ke_vs_dil_all3}
 \end{subfigure}\\
 \begin{subfigure}[b]{0.49\textwidth}
  \centering
  \subcaption{}
     \input{./diss_vs_vort_all3}
     \label{fig:diss_vs_vort_all3}
 \end{subfigure}
 ~\hspace{0.02\textwidth}
 \begin{subfigure}[b]{0.49\textwidth}
  \centering
  \subcaption{}
     \input{./ke_vs_vort_all3}
     \label{fig:ke_vs_vort_all3}
 \end{subfigure} 
 \caption{4D representation of the family of solutions across Mach number, ranging from 0.75 to 0.8.}
\label{fig:phase_portrait_Re_2000_all3}
\end{figure}

\subsection{High Mach number regime}

The trajectories corresponding to the highest Mach numbers ($0.75<M<0.8$) from the family of solutions shown in figure \ref{fig:Mach_range} are gathered in figure \ref{fig:phase_portrait_Re_2000_all3}. In this regime, the high flow compressibility alongside the phase of the flow-acoustic interaction favour the appearance of a new acoustic wave caused by the dipole associated to the shear layer vortex. This new acoustic wave will be referred to as the shear layer acoustic wave. As mentioned earlier, the acoustic radiation from the trailing edge arises from the impingement of both low and high density flow events at this location. The low density events are related to the shear layer vortex which merges with the stationary vortex, whereas the high density ones result from the counter-rotation of the stationary and shear layer vortices which compresses the flow amongst them. For the lowermost Mach number orbits ($M<0.65$), the higher speed of sound relative to the flow velocity results in an favourable (in phase) synchronisation of the shear layer and trailing edge dipoles, which further enhance the upstream radiated acoustic wave. At the intermediate Mach number regime, the propagating velocity of sound is less and the shear layer dipole grows in magnitude. These two phenomena partially cancel out the upstream sound radiation (out of phase synchronisation), which shifts the leading sound radiation towards the vertical. At the current Mach number range, the large flow compressibility has reinforced the shear layer dipole as a leading contributor in the overall sound radiation of the system. In addition, the speed of sound continues to reduce which increases the lag of the interaction between the acoustic wave radiated from the trailing edge and the shear layer. Recalling figure \ref{fig:dil_vs_M}, the time-averaged norm of dilatation experiences a sudden drop after $M=0.76$, reaching a local minimum at $M=0.78$ and finally rapidly increasing again until the upper end of out Mach number range. This non-monotonic behaviour differs from the orbits at the lower Mach numbers and it is also reflected in figures \ref{fig:diss_vs_dil_all3} and \ref{fig:ke_vs_dil_all3}. In these figures we observe that the oscillations in the averaged norm of dilatation are not caused by a single event with a radically higher or lower compressibility, but by the entire orbit which raises and lowers the flow compression as a whole. The origin of this global changes resides in both the increasing flow compressibility as the Mach number is raised and the phase of the interaction between the acoustics and the shear layer dipole. Note that this interaction also occurs for the lower Mach numbers, but there, due to the weaker shear layer dipole, these changes in dilatation were masked by the constantly increasing flow compressibility with the Mach number. 

\begin{figure}
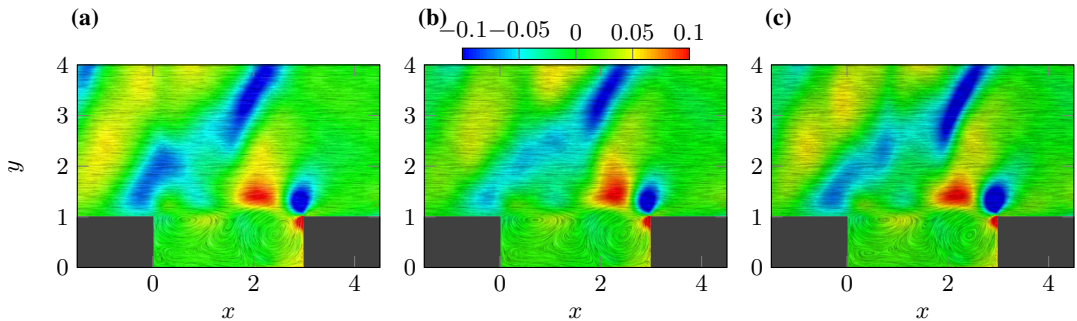

 \begin{subfigure}[b]{0.32\textwidth}
  \centering
  \subcaption{}
     \input{./dil_max_vort_M076}
     \label{fig:dil_max_vort_M076}
 \end{subfigure}
 ~\hspace{0.00\textwidth}
 \begin{subfigure}[b]{0.32\textwidth}
  \centering
  \subcaption{}
     \input{./dil_max_vort_M078}
     \label{fig:dil_max_vort_M078}
 \end{subfigure}
 ~\hspace{0.00\textwidth}
 \begin{subfigure}[b]{0.32\textwidth}
  \centering
  \subcaption{}
     \input{./dil_max_vort_M080}
     \label{fig:dil_max_vort_M080} 
 \end{subfigure}\vspace{-4.15cm}
 \begin{subfigure}[b]{\textwidth}
 	\centering
 	\hspace{-0.15cm}
  	\input{./dil_max_vort_colourbar}
 \end{subfigure}
 \vspace{3.0cm}
 \caption{Instantaneous contours of the dilatation field at Mach numbers 0.76 \textbf{(a)}, 0.78 \textbf{(b)} and 0.8 \textbf{(c)}. These instants correspond to the maximum norm of the vorticity field along the periodic orbit.}
\label{fig:dil_max_vort}
\end{figure}

Similarly, all quantities with the exception of the norm of kinetic energy do not follow a constant trend but instead oscillate. Contrarily, the overall kinetic energy keeps increasing as the Mach number is raised, but this time the phenomenon has nothing to do with the changing flow compressibility. As we will see, the cause of this phenomenon is the monotonic drop in $\Theta$ seen in figure \ref{fig:Re_thet_vs_M}. For a more intuitive interpretation of these physical events, figure \ref{fig:dil_max_vort} shows the snapshots of the dilatation field at Mach numbers 0.76, 0.78 and 0.8 at the point of maximum norm of the vorticity field (figure \ref{fig:phase_portrait_Re_2000_all3}). The position of these snapshots are highlighted in figure \ref{fig:ke_vs_vort_all3} with red square symbols at the flow orbit and it corresponds to the instant when the vortex impingement (low density event) occurs. In the same figure, the red circles show the opposite phase to the plots from \ref{fig:dil_max_vort}, which indicates the impingement of the high density area onto the trailing edge. The interference amongst the upstream propagating acoustic wave and shear layer acoustic wave, which we observed for the first time earlier on in figure \ref{fig:dil_74_max}, grows continuously through $M=0.76$ (figures \ref{fig:dil_max_vort_M076}), reaching the anti-phase of the interaction at $M=0.78$. At this point, both acoustic waves cancel each other out in the vicinity of the cavity (figure \ref{fig:dil_max_vort_M078}). This phenomenon causes the sudden drop in the average dilatation field from $M=0.76$ to $M=0.78$. From this Mach number onwards, the shear layer becomes more energetic, and the phase of the interaction slowly becomes favourable again. Note that the features of the purely convecting events (i.e.~vortices) are barely altered throughout this interval. Additionally, despite the considerable difference in Mach number, the shape and alignment of the dipoles related to convective phenomena are remarkably similar to the ones illustrated in figure \ref{fig:cartoon_a} for the $M=0.5$ case. Furthermore, the qualitative agreement of figure \ref{fig:dil_max_vort_M080} with the data shown by \cite{rowley2002self} (their figure 6) is excellent.

\subsection{Momentum thickness effect and stability}
\label{subsec:mom_t_and_s}

In the family of periodic orbits presented above, the optimisation algorithm induced a progressive drop in the incoming boundary layer as the Mach number was raised (figure \ref{fig:Re_thet_vs_M}). In order to isolate the effect of this phenomenon over the periodic orbits, we have computed an additional periodic solution at $M=0.8$ with the same incoming boundary layer as the initial orbit at $M=0.5$. The $Re_\Theta$ of this new orbit is 64.37, whereas the continued orbit at the same Mach number presents a $Re_\Theta=23.06$. After a preliminary observation to these orbits \citep{mythesis}, we have concluded that the orbits describe trajectories of identical shape, where the new periodic solution presents a negative shift on its average norms of dilatation and kinetic energy. The fact that the shape in both trajectories is the same it suggests that the physical mechanisms which govern the flow remain unchanged. In addition, the period of the new orbit is almost identical to the one from the respective continued orbit (see appendix \ref{appA}), which confirms that both sets of solutions correspond to the same Rossiter mode. On the other hand, the thicker incoming boundary layer in the new orbit is the physical event that is causing the negative shift in dilatation, but mainly also in kinetic energy. Bear in mind that this phenomenon was also observed in figure \ref{fig:ke_vs_vort_all3}, where all the periodic solutions from this particular Mach number range exhibit a progressive shift in the kinetic energy norm as the boundary layer thickness is reduced. One of the primary effects of this smaller $Re_\Theta$ in the continued flow trajectory (caused only by the decrease of $\Theta$) is a more unstable character of the shear layer \citep{brs2008}. This enhanced instability yields a stronger leading edge vortex, which following the flow mechanisms described earlier in section \ref{sec:PS_M050}, it eventually travels downstream and impinges onto the cavity's trailing edge, radiating a stronger acoustic wave.

In order to assess the stability of the periodic solutions presented in this chapter, we introduced random noise perturbations across the entire flow-field for several flow orbits. Despite the amplitude of these disturbances ranging up to $10\%$ of $Q$ in some situations, the flow-field eventually adjusted back to the unperturbed trajectory in every case, stabilised by the flow-acoustic feedback mechanism. In addition, bear in mind that random noise perturbations are essentially strong numerical point-to-point oscillations in the flow-field. For this reason, they might be interpreted as spurious oscillations by the high-order explicit filter applied by the DNS code \citep{rsand}, which partially removes these disturbances, contributing to the stability of the periodic orbits. Instead, to more efficiently perturb the periodic solutions, we now introduce an initial body force disturbance in the streamwise and vertical momentum state variables. The spatial domain of activity for this perturbation is defined as
\begin{equation}
W_f \left(\vec{x}\right) = \mathrm{e}^{-\frac{\left(x-x_0\right)^2+\left(y-y_0\right)^2}{0.05}},
\end{equation}
where both streamwise and vertical momentum components experience an external forcing up to $-0.1\rho_\infty u_\infty$ at the centre of the Gaussian function, at $\vec{x}_0=\left(1,1\right)$. The exact location of this forcing has been carefully chosen to alter the flow-field at the shear layer and the proximity of the cavity's leading edge. If any exists, an unstable shear layer mode would exhibit its highest receptivity values at these particular spatial regions (see section \ref{subsec:stability_of_equilibria}). The flow trajectories we perturb are the flow solution at $M=0.8$ continued from the original periodic trajectory at $M=0.5$ (named M080), and the one obtained straight from the developed flow (M080-fd). The flow trajectories representing perturbed orbits are appended with the prefix `p-' (i.e. p-M080 for the perturbed M080 and so on).

\begin{figure}
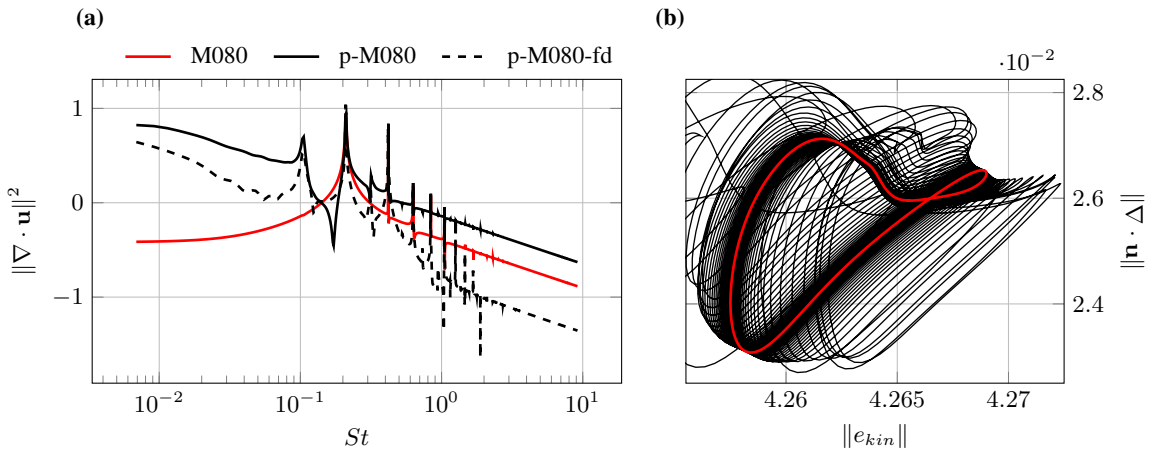

 \begin{subfigure}[b]{0.55\textwidth}
  	\centering
  	\subcaption{}
  	\vspace{-0.175cm}
    \input{./stability_M080_fft}
    \label{fig:stability_M080_fft}
 \end{subfigure}
 ~\hspace{0.0\textwidth}
 \begin{subfigure}[b]{0.43\textwidth}
	\centering
  	\subcaption{}
    \input{./stability_M080_zoom}
    \label{fig:stability_M080_zoom}
 \end{subfigure}
 \caption{\textbf{(a)} Frequency spectrum of the perturbed flow trajectories at $M=0.8$ and the exact (continued) solution also at $M=0.8$. \textbf{(b)} Final stage of p-M080 decaying to M080.}
 \label{fig:stability_M080_more}
\end{figure}

Figure \ref{fig:stability_M080_fft} displays the Fourier transformed norms of the dilatation field across time from M080, p-M080 and p-M080-fd. For simplicity, M080-fd has not been represented as it yields almost identical results to M080. Indeed, due to the high receptivity of the area surrounding the cavity's leading edge, this perturbation initially triggers the first Rossiter mode for both p-M080 and p-M080-fd, which appears as a peak at $St\approx 0.103$. Taking now a closer look on how M080-p decays to M080 (figure \ref{fig:stability_M080_zoom}), we can spot the influence of the first Rossiter mode on the trajectory described by p-M080. Once the perturbed state p-M080 reaches the vicinity of M080, it commences an oscillation about the M080 flow solution with both frequency and exponential decaying rate associated with this stable first Rossiter mode. The stability of these Rossiter modes is related to the mode selection phenomenon also discussed by \cite{brs2008}. This flow mechanism consists of the cavity flow `choosing' one Rossiter mode to govern the shear layer dynamics, where all the remaining modes below experience an exponential decay. This mode selection is not fully understood yet, and according to the literature, it appears to be dependent on parameters such as Reynolds number or the cavity's aspect ratio. \cite{brs2008} reported for their 2M06 case that the shear layer oscillated with the first Rossiter mode. Similarly to the present investigation, in the initial stages of time marching, they also observed an additional Rossiter mode (in their case the second mode) with relevant activity, which eventually decayed completely. Furthermore, on a potential three-dimensional scenario, they suggested that the interaction with the spanwise modes might affect the selection of the dominating Rossiter mode. 

\subsection{Overall Sound and Directivity}
\label{subsec:oaspl_dir}

\begin{figure}
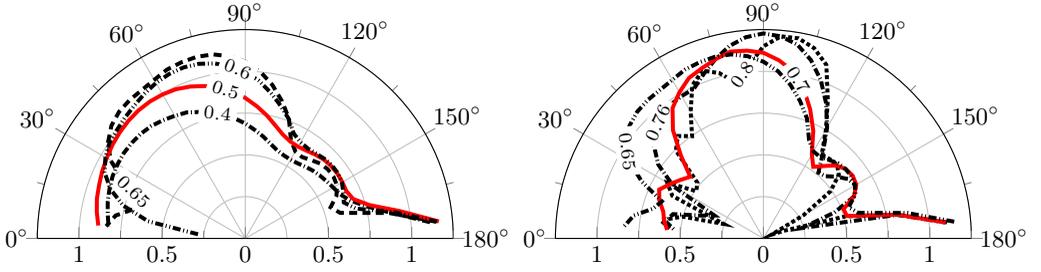

\begin{subfigure}[b]{0.49\textwidth}
  	\centering
  	\subcaption{}
    \input{./oaspl_polar}
    \label{fig:oaspl_polar1}
 \end{subfigure}
 ~\hspace{0.0\textwidth}
 \begin{subfigure}[b]{0.49\textwidth}
  	\centering
  	\subcaption{}
    \input{./oaspl_polar2}
    \label{fig:oaspl_polar2}
 \end{subfigure}
 \caption{OSPL directivity across Mach number.}
\label{fig:oaspl_polar}
\end{figure}

So far in the present section, we have seen how the overall behaviour of the periodic orbits changes as a consequence of the phase modification of the interaction amongst the shear layer and the trailing edge acoustic radiation. In particular, for the higher Mach numbers $\left(M>0.65\right)$ the sound radiation of the shear layer is of comparable magnitude to the radiation from the trailing edge. This leads to an enhanced interaction which, as seen in the above dilatation plots, has an effect on the energy and directivity of the overall sound radiation. To characterise the sound (or noise) radiation for the current family of periodic orbits we define the overall sound pressure level as \begin{equation}
    \mathrm{OSPL} = 10 \cdot log \left( \frac{\int^{\infty}_\infty \left| \mathcal{F} \left( p'\left(t\right) \right) \right|^2 \mathrm{d}f}{\left(p_{ref}\right)^2}\right),
\label{eq:ospl}
\end{equation}
where $p'\left(t\right)$ are the pressure fluctuations at the measurement point, $p_{ref}$ is the reference pressure level set as $2 \cdot 10^{-5}$ and $\mathcal{F}$ indicates a Fourier transform. To evaluate the sound directivity, the OSPL was computed in 35 equally distributed monitor points along an arc of radius 5, each of them separated by 5 degrees. The centre of the arc is located at coordinates 1.5 and 1 in the streamwise and vertical directions, respectively. The OSPL of the most representative periodic orbits are gathered in figure \ref{fig:oaspl_polar}, where angles lower than 90 degrees and greater than 90 degrees show, respectively, an upstream and downstream sound radiation. The OSPL values are normalised with the maximum upstream propagating OSPL value from the orbit at $M=0.5$. 

In section \ref{sec:PS_M050} we observed how the shear layer and leading edge dipoles enhance the acoustic wave originally radiated by the trailing edge at $M=0.5$. At the same time, the interaction of the shear layer and trailing edge dipoles partially cancels out the sound radiation in the downstream direction (figure \ref{fig:cartoon_a}). This behaviour agrees with the content of figure \ref{fig:oaspl_polar1}, where the OSPL values reach higher magnitudes at the upstream propagating angles, and the downstream propagation is severely mitigated by the shear layer dipole. The quasi-circular shape of the plots at Mach numbers 0.4 and 0.5 indicate that this upstream sound radiation is almost entirely governed by the cavity's trailing edge noise. Note that this is not the case for $M=0.6$ and  $M=0.65$, where the shear layer is compressible enough to start radiating sound. As the Mach number is further increased, the higher flow compressibility leads to stronger dipoles at the leading edge and shear layer. This phenomenon in combination with a lower speed of sound forces the above-mentioned interaction to slowly get out of synchronisation. The OSPL values for Mach numbers 0.6 and 0.65 show how the dipole located at the leading edge cancels out the upstream sound propagation in the vicinity of the wall. Furthermore, the high intensity of the shear layer dipole is reflected in the more pronounced lobe at about $140^{\circ}$. Note that as the Mach number is increased, the leading sound radiation direction shifts monotonically towards a higher angle. This trend also applies for the higher Mach numbers (figure \ref{fig:oaspl_polar2}). At Mach numbers from 0.65 to 0.70, the dipole interaction continues out of synchronisation which leads to a substantial reduction of the acoustic radiation in the upstream direction. As the interaction becomes favourable again from $M=0.70$ to $M=0.76$, the upstream sound radiation increases in the $30^{\circ}$ to $60^{\circ}$ range. This is solely caused by the higher radiation of the shear layer dipole at this Mach number range. As shown in figure \ref{fig:dil_max_vort}, for Mach numbers above $M=0.76$ the sound radiation is split in two different waves, where now the shear layer dipole is the responsible from the upstream propagating acoustics. This is observed in figure \ref{fig:oaspl_polar2} as two main distinct lobes for $M>0.76$. The sound radiation from the cavity's trailing edge is cancelled out almost completely in the upstream direction as it interacts with the shear layer acoustics, which forces its propagation in the vertical direction.

\section{Equilibrium Solutions}
\label{sec:equilibria}

\begin{figure}
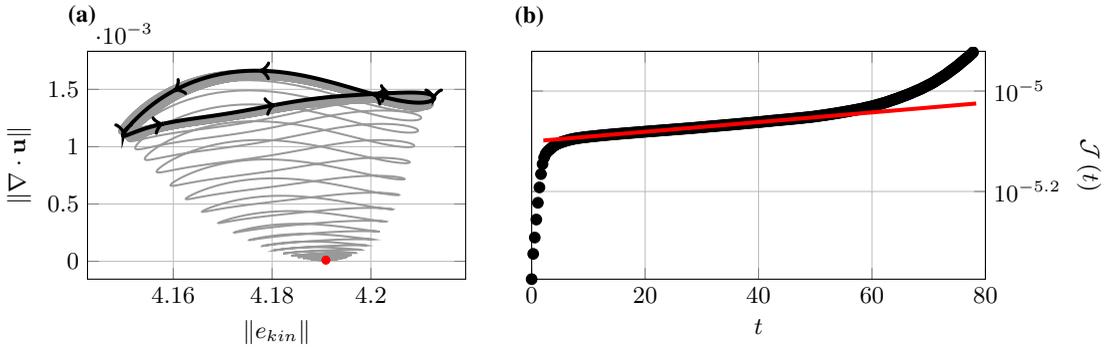

	\begin{subfigure}[b]{0.42\textwidth}
  		\centering
  		\subcaption{}
  		\vspace{-0.3cm}
    	\input{./steady_to_periodic_M050}
    	\label{fig:steady_to_periodic_transition}
	\end{subfigure}
 	~\hspace{0.0\textwidth}
 	\begin{subfigure}[b]{0.55\textwidth}
  		\centering
  		\subcaption{}\hspace{-0.90cm}
  		\input{./transition_cost_vs_time}
  		\label{fig:transition_cost}
  	\end{subfigure}
 \caption{\textbf{(a)} Transition from the steady solution to its associated periodic orbit at $M=0.5$. \textbf{(b)} Evolution of the cost function in the initial stage of the transition from steady to periodic flow.}
\label{fig:M050_steady_to_periodic}
\end{figure}

In the preceding section we have shown how the periodic solutions are dominated by compressible events. As the Mach number was decreased down to the incompressible regime ($M \approx 0.30$), the amplitude of this periodic behaviour decayed to zero, giving rise to a steady state. Hence, as shown in figure \ref{fig:Mach_range}, this incompressible limit could be seen as a bifurcation point amongst the families of periodic and steady or equilibrium solutions. In the sequel we show that, away from the lower Mach number regime ($M\gtrsim 0.35$), these steady solutions are unstable, where a perturbation suffices to trigger a transition towards the limit cycle (figure \ref{fig:steady_to_periodic_transition}). Bear in mind that the steady solutions are not seen in a naturally evolving flow at compressible Mach numbers (where the compressible effects are not negligible $M \gtrsim 0.30$). Interestingly, the flow topology of these equilibrium states matches the flow features shown earlier in figure \ref{fig:M025_vort}. Similarly to the periodic orbits, these solutions consist of a weak shear layer vortex which counter-rotates with respect to the stationary vortex, located at the downstream end of the cavity. From the same figure, we also see that the stationary and shear layer vortices interact to maintain a fragile equilibrium. A small perturbation crossing the shear layer would trigger the merging process between these two vortices as we saw in section \ref{sec:PS_M050}. These particular flow features remain almost unaltered throughout the entire Mach number range, where only the norms of dilatation (figure \ref{fig:dil_vs_M}) and kinetic energy (figure \ref{fig:ke_vs_M}) show significant changes. The flow variations reflected in these two norms are strongly linked to the higher flow compressibility associated to a higher Mach number. Similarly to the periodic orbits, the counter-rotating character of the shear layer and stationary vortices gives rise to a flow compression at the mid-point of the two vortices. Just downstream from this location, the stationary vortex further accelerates the flow on top of the shear layer, which causes a flow expansion. However, the strongest compressible phenomenon in this family of equilibrium solutions is the flow compression in the vicinity of the trailing edge of the cavity, which emanates from the constant flow impingement (this time steady impingement) onto the trailing edge. Thus, the increase in the norm of dilatation is caused by the strongest density gradients in the flow, whereas the increase in kinetic energy arises mainly from the stronger compressible effects above the stationary vortex.

\subsection{Stability of Equilibrium Solutions}
\label{subsec:stability_of_equilibria}

The bifurcation amongst the families of steady and periodic solutions was illustrated in figure \ref{fig:Mach_range}. In this figure, it is straightforward to notice how the periodic and steady families get progressively closer as we descend in Mach number and completely merge for Mach number $M\leq0.3$. Thus, in order to determine the nature of the bifurcation, we require a detailed stability analysis of these equilibrium solutions. To this end, we make use of both forward (non-linear) and adjoint (linear) Navier-Stokes solvers available in our code \citep[see][for a description of the adjoint framework used]{mythesis}. Respectively, forward and adjoint stability analysis yield information about the system's response and receptivity, where they can also be combined to reveal the so-called `{wave maker}' or core of the instability. For an elaborated introduction to global stability analysis and the use of forward and adjoint global modes, the reader is referred to the reviews by \cite{vassilios_gls} and  \cite{luchini_adjoint}. To perform the stability analysis, we use dynamic mode decomposition (DMD) to get the eigenvalues and eigenvectors (modes) of the system straight from the data produced by our numerical solvers, without any further modifications. This approach seeks a flow decomposition of the form \begin{equation}
\tilde{Q}\left(t\right) = Q_0 + \sum_{n=1}^{N_{mod}} \Psi_n e^{\mu_n t} + c.c., 
\label{eq_2:gm_decomposition}
\end{equation} where $c.c.$ indicates complex conjugate and $\Psi_n$ and $\mu_n$ are the system's eigenvectors and eigenvalues. According to \cite{dmd}, DMD recovers the global modes of a linear process, whereas for the non-linear case ``it identifies the dominant frequencies and their associated spatial structures''. Since we lack a linearised Navier-Stokes solver, the forward global modes are approximated assuming linear behaviour of the full non-linear Navier-Stokes equations in the initial stages of time marching. Note that for unstable steady solutions, the flow undergoes a transition from a steady state to periodic orbit. Consequently, extending the dataset too much might result in the DMD framework accounting for non-linear dynamics related to the limit cycle, whereas a short dataset might not produce converged results. On the other hand, the adjoint equations are linear, which allows extending the adjoint dataset as required to obtain converged adjoint dynamic modes. Hence, since the spectrum recovered from the forward and adjoint simulations should coincide, we use the adjoint eigenvalues as the reference and also to check the approximated forward modes.
 
\subsubsection{Global Stability Analysis at $M=0.5$}

From figure \ref{fig:M050_steady_to_periodic} we can see that the equilibrium solution at $M=0.5$ is unstable. Figure \ref{fig:steady_to_periodic_transition} shows the evolution of this unstable solution, where shortly after the beginning of the simulation, the flow instability experiences low amplitude oscillations which grow exponentially in time. This amplitude growth continues until the flow-field approaches the vicinity of the periodic solution. There, the amplitude's growth rate decays rapidly, where the flow saturates and follows closely the trajectory described by the periodic orbit. Therefore, in order to approximate the forward global modes using DMD, we must establish a time interval where the global behaviour of the system can be approximated as linear. Figure \ref{fig:transition_cost}, shows the value of the cost function (\ref{eq:steady_cost_f}) over time. The initial steep increase in the cost function is the result of the flow solver reacting to an initial condition which was computed externally. For instance, the boundary conditions and derivative routines adapt the steady solution to satisfy the physical conditions imposed at the boundaries, and also fit the flow-field onto a polynomial function of the same order as the derivative routines. After this initial `rejection' is overcome, one would expect that the divergence of the flow-field from the initial steady solution would be much more drastic once the non-linear effects start taking place. We estimate the start of the non-linear flow behaviour from figure \ref{fig:transition_cost}, as the point where the trajectory of the cost function abandons the linear trend exhibited approximately from 12 to 47 time units. On the other hand, the converged eigenvalue spectrum obtained through the adjoint simulations used 801 snapshots sampled equidistantly across a time span of 220 time units. To avoid unwanted influence from the boundary conditions, the snapshot sequences (both adjoint and forward datasets) were cropped down to the vicinity of the cavity. 

\begin{figure}
	\centering
    \input{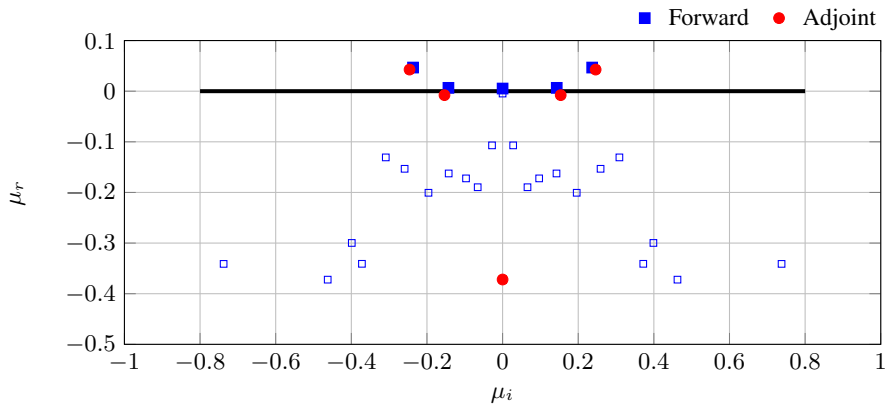}
    \caption{Eigenvalue spectrum for the steady solution at $M=0.50$. The red circular markers show the eigenvalues calculated through the adjoint simulations. The blue squared symbols represent the approximate eigenvalues obtained using the non-linear forward simulations. The smaller squared symbols show purely numerical modes (noise), which is a product of the insufficiently large forward snapshot sequence.}
 	\label{fig:M050_spectrum}
\end{figure}

Figure \ref{fig:M050_spectrum} shows the eigenspectrum calculated using both forward and adjoint snapshot sequences. These eigenvalues have been transformed to the continuous-time form using the relations $\mu_r = \operatorname{\mathbb{R}e}\lbrace log\left(\lambda_n\right)\rbrace/\Delta t$ and $\mu_i = \operatorname{\mathbb{I}m}\lbrace log\left(\lambda_n\right)\rbrace/\left(2 \pi \Delta t\right)$, indicating growth rate and Strouhal number ($St = fD/U_\infty$), respectively. The resulting leading (most unstable) eigenvalues yielded by both forward and adjoint approaches are very similar. The exact values produced by the adjoint DMD for this particular mode are $\mu_r=0.042644$ and $\mu_i=0.246062$, where the forward DMD results are just $3.33\%$ off. Also note the closeness of both approaches on the secondary eigenvalue at $\mu_r=-0.007837$ and $\mu_i=0.153670$. The remaining unmatched eigenvalues predicted by the forward DMD appear as a direct consequence of an insufficiently long snapshot sequence to produce a fully converged spectrum. This phenomenon was also observed when studying the convergence of the adjoint spectrum, where most of the damped eigenvalues ($\mu_r < 0 $) further decayed as the snapshot sequence was incremented. These stable eigenvalues (with  $\mu_r < -0.5 $) have been excluded from the above plot due to their lack of dynamical relevance. Hence, figure \ref{fig:M050_spectrum} can be seen as a validation of both forward and adjoint compressible Navier-Stokes solvers, alongside the DMD post-processing tool. 

\begin{figure}
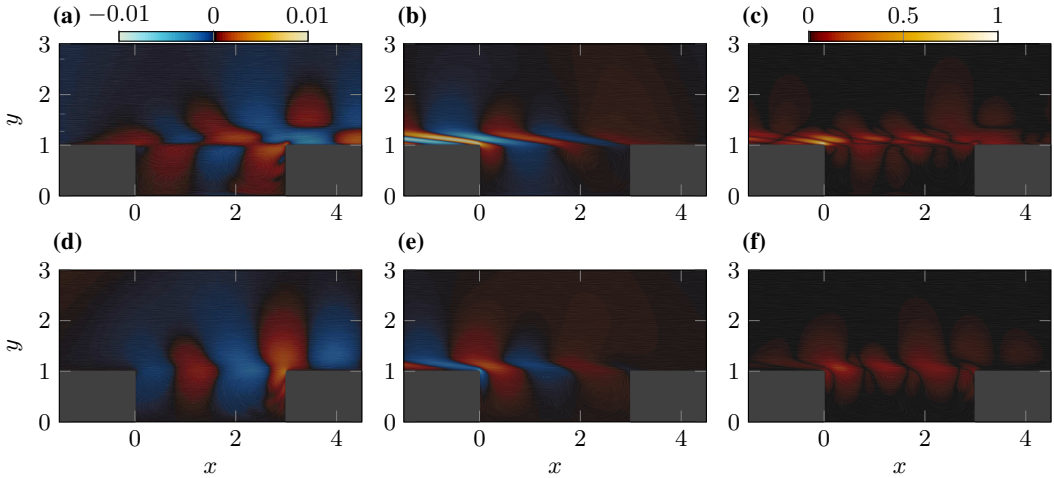

 \newlength\spacingfigmodeone
 \setlength\spacingfigmodeone{-0.25cm}
 \newlength\spacingfigmodetwo
 \setlength\spacingfigmodetwo{-0.00cm}
 \begin{subfigure}[b]{0.32\textwidth}
  	\centering
  	\subcaption{}
    \vspace{\spacingfigmodeone}
    \input{./M050_forward_mu}
    \label{fig:M050_a}
 \end{subfigure}
 ~\hspace{0.00\textwidth}
 \begin{subfigure}[b]{0.32\textwidth}
  	\centering
  	\subcaption{}
    \vspace{\spacingfigmodeone}
    \input{./M050_adj_mu}
    \label{fig:M050_b}
 \end{subfigure}
 ~\hspace{0.00\textwidth}
 \begin{subfigure}[b]{0.32\textwidth}
  	\centering
  	\subcaption{}
    \vspace{\spacingfigmodeone}
    \input{./M050_u_wavemaker}
    \label{fig:M050_c}
 \end{subfigure}\\
 \begin{subfigure}[b]{0.32\textwidth}
  	\centering
  	\subcaption{}
    \vspace{\spacingfigmodeone}
    \input{./M050_forward_mv}
    \label{fig:M050_d}
 \end{subfigure}
 ~\hspace{0.00\textwidth}
 \begin{subfigure}[b]{0.32\textwidth}
 	\centering
 	\subcaption{}
    \vspace{\spacingfigmodeone}
    \input{./M050_adj_mv}
    \label{fig:M050_e}
 \end{subfigure}
 ~\hspace{0.00\textwidth}
 \begin{subfigure}[b]{0.32\textwidth}
 	\centering
  	\subcaption{}
    \vspace{\spacingfigmodeone}
    \input{./M050_v_wavemaker}
    \label{fig:M050_f}
 \end{subfigure}\vspace{-6.4cm}\\
 \begin{subfigure}[b]{0.32\textwidth}
  	\centering
    \input{./M050_fwd_adj_mode1_colourbar}
 \end{subfigure}\hspace{0.35\textwidth}
  \begin{subfigure}[b]{0.32\textwidth}
  	\centering
    \input{./M050_wavemaker_colourbar}
 \end{subfigure}\vspace{6.0cm}
\caption{Streamwise (top) and vertical (bottom) components of the forward $\left( \textbf{(a)}, \textbf{(b)}\right)$ and adjoint $\left( \textbf{(b)}, \textbf{(c)}\right)$ momentum components associated with the leading eigenvalue for the equilibrium solution at $M=0.5$. Figures \textbf{(c)} and \textbf{(f)} show the structural sensitivity maps of the streamwise and vertical momentum components, calculated as $\big|\Psi^*_{m_u^*}\big| \cdot \big|\Psi_{\rho u}\big|$ and $\big|\Psi^*_{m_v^*}\big| \cdot \big|\Psi_{\rho v}\big|$, respectively. The contour levels of these two plots are normalised with $6\cdot10^{-6}$.}
\label{fig:M050_modes_and_wavemaker}
\end{figure}

Figure \ref{fig:M050_modes_and_wavemaker} presents the forward and adjoint modes, and the wave maker corresponding to the momentum components of the most unstable eigenvalue. Since the current equilibrium solution is unstable, the forward mode reveals the laminar shear layer and the reattached laminar boundary layer as the leading amplifiers of this instability. Apart from the amplitude, the locations where the mode switches sign reveal the underlying physical mechanisms which drive the dynamics. In substance, loci with contrary sign represent opposite flow behaviour, such as increase-decrease of pressure, momentum, density, etc. For reference, the streamlines calculated from the velocity field of the steady flow solution are superimposed on the contour plots in figure \ref{fig:M050_modes_and_wavemaker}. Curiously, the momentum mode shapes highlight the close relationship between the vortices located in the shear layer and also at the downstream end of the cavity. If we analyse in detail the streamwise momentum component, we observe that it switches signs in the proximity of the vertical centreline of the vortices located in the shear layer and at the downstream end of the cavity. This phenomenon suggests that both vortices are compressed and expanded in the streamwise direction at the eigenvalue's frequency. Further, as also seen in figure \ref{fig:M050_a}, similar sign changes occur across the laminar shear layer in the vertical direction, revealing its unstable character. Additionally, the vertical momentum component also shows relevant activity along the laminar shear layer. Figure \ref{fig:M050_d} uncovers patches of opposite increase in vertical momentum, where one of the sign changes takes place just in between the above-mentioned shear layer and downstream vortices. Hence, the combination of the streamwise and vertical momentum components induces fluctuations in the laminar shear layer which affect both shear layer and downstream vortices. These two vortices begin an oscillating motion with increasing amplitude, resulting in the downstream vortex absorbing the shear layer vortex. As detailed in previous sections, the extra streamwise momentum resulting from this merging process impinges onto the cavity's trailing edge, which radiates upstream an acoustic wave, initiating the Rossiter mode. From this point onwards, the system can no longer be approximated as linear and begins the last stage of its transition towards the stable limit cycle (figure \ref{fig:steady_to_periodic_transition}).

\begin{figure}
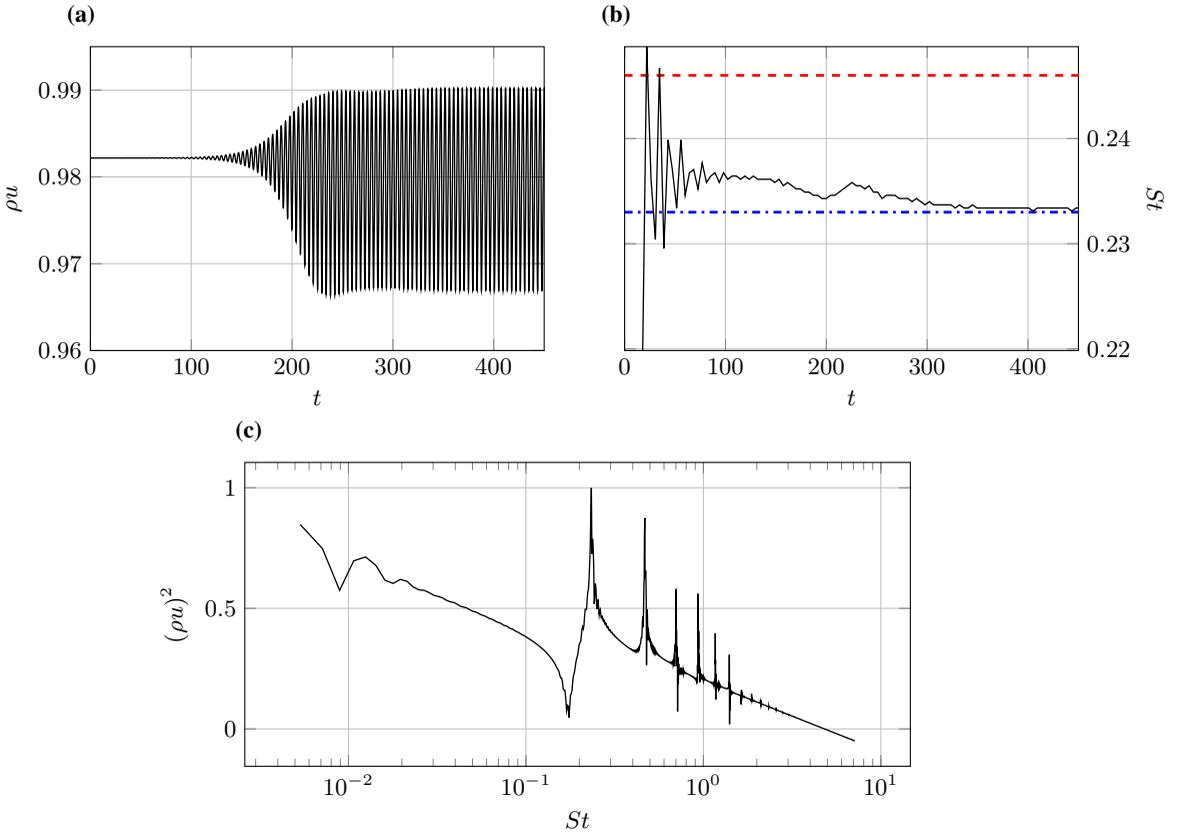

 \begin{subfigure}[b]{0.49\textwidth}
  \centering
  \subcaption{}
     \input{./U_monitor_transition_M050}
     \label{fig:U_monitor}
 \end{subfigure}
 ~\hspace{0.02\textwidth}
 \begin{subfigure}[b]{0.49\textwidth}
  \centering
  \subcaption{}
     \input{./frequency_evolution_M050}
     \label{fig:freq_evolution_M050}
 \end{subfigure}\\
 \begin{center}
  \begin{subfigure}[b]{0.67\textwidth}
  \centering
  \subcaption{}
     \input{./transition_M050_fft}
     \label{fig:transition_M050_fft}
 \end{subfigure}
 \end{center}
 \caption{\textbf{(a)} Time signal of the streamwise momentum at $\vec{x}=\left(2.25,1\right)$. \textbf{(b)} Estimated flow-acoustic feedback loop's frequency evolution across time. Red dashed and blue dash-dotted horizontal lines indicate the frequencies associated the unstable mode and periodic limit cycle, respectively.  \textbf{(c)} Power spectrum of the time signal show in \textbf{(a)}.}
 \label{fig:Transition_M050}
\end{figure}

So far, we know the physical mechanisms associated with the unstable eigenvalue, which trigger a transition from a steady state towards a periodic orbit in the present 2D cavity flow. In other words, we only know that our steady flow solution is unstable and how it reacts to that instability. In order to understand the origin of the instability, we recover the adjoint mode of that unstable eigenvalue. The adjoint modes can be interpreted as receptivity maps which indicate how to most efficiently trigger their respective forward mode. Hence, the adjoint mode shown in figures \ref{fig:M050_b} and \ref{fig:M050_e} reveals the receptivity of its forward mode. Similarly to the forward mode, the streamwise and vertical adjoint momentum components (figures \ref{fig:M050_b} and \ref{fig:M050_e}, respectively) present a considerable activity. Opposite to the behaviour exhibited by the forward mode, the adjoint mode shows increasing amplitude in the upstream direction, highlighting the vicinity of the leading edge as particularly receptive. This high receptivity, linked to the separated shear layer and the incoming boundary layer, is responsible for the global instability. Any perturbation in the state vector where its corresponding adjoint mode component is not zero will set off the instability growth. For the current equilibrium solution, the numerics from our compressible Navier-Stokes solver, alongside with the numerical precision of the flow solution, suffice to trigger this unstable mode. Following the concept of structural sensitivity (or wave maker), we can now combine the forward and adjoint modes. Such analysis reveals the area of the flow-field which acts as `the driver of the oscillation' \citep{luchini_adjoint}, highlighting the regions with both high receptivity and response. These spatial maps are also referred to as sensitivity to a localised feedback \citep{luchini_2007}. Figures \ref{fig:M050_c} and \ref{fig:M050_f} show the wave maker regions for the streamwise and vertical momentum components. These structural sensitivity maps expose predominately both the shear layer and the incoming laminar boundary layer as the origin of the instability. Especially, the areas showing a greater intensity are located in the proximity of the cavity's leading edge, where the adjoint mode peaks. In addition, as observed in figure \ref{fig:M050_c}, the laminar boundary layer only contributes to the onset of the instability in the streamwise direction. In fact, the point of highest magnitude is located just above the leading edge, before this incoming laminar boundary layer separates from the wall. Moreover, the high structural sensitivity patches exhibited along the shear layer in figure \ref{fig:M050_f} are also strongly linked to the flow structures located underneath. The two weakest patches (yet non-negligible) are located in the shear layer, precisely above the cores from the downstream and shear layer vortices. Overall, for this distinct geometry, the wave maker region is distributed along the incoming boundary layer and shear layer, peaking in intensity at the cavity's leading edge, to then decay towards the trailing edge. In particular, the activity shown in these two wave maker components vanishes rapidly after the cavity's trailing edge, because the convective-acoustic feedback mechanism does not hold past the trailing edge. Analogously to the acoustic-convective feedback loop occurring in periodic cavity solutions, any minimal flow imbalance which encounters the trailing edge will reflect upstream a weak acoustic wave. This upstream travelling acoustic wave eventually reaches the proximity of the leading edge of the cavity, setting in motion the unstable mode. Hence, once the unstable mode is active, it will further disturb the flow-field producing a more energetic acoustic wave than the precedent in the following impingement on the trailing edge. This feedback loop continues fuelling the flow instability until the linear flow dynamics collapse. Once this occurs, the natural frequency from the flow-acoustic feedback loop progressively evolves from the frequency of the (linear) unstable mode of $St\approx0.246$, to the characteristic frequency of its corresponding limit cycle (Rossiter mode) of $St\approx0.233$. A similar disagreement amongst the frequencies associated with the unstable eigenmode and the non-linear periodic solution was also observed by \cite{sipp2007global} in an incompressible flow over a square cavity. Figure \ref{fig:U_monitor} shows the streamwise momentum time signal through this transition, captured by a monitor point located at the shear layer. At this location, the unstable forward mode is considerably active, which permits the estimation of the dominant frequency by tracking the local minima (or maxima) of this signal across time. Since there is only one single unstable mode, the frequency should asymptotically approach the frequency of this unstable eigenvalue, which governs the long time behaviour of the system. Figure \ref{fig:freq_evolution_M050} shows his frequency estimation as a function of time, where it appears to become stable after approximately 100 time units. The fact that the initial strong frequency oscillations ceased is an indication that all the stable modes have decayed significantly, leaving the unstable mode as the only one active in the system. At this point in time, we readily observe that the approximated frequency has already abandoned the frequency associated with the unstable mode, which confirms that the dynamics have stopped being governed by linear flow mechanisms \citep{devicente2014}. Furthermore, consider that the linear behaviour ceases even before the vortex merging of the shear layer and downstream vortices begins, which occurs approximately after 135 time units. After this, the flow undergoes the last stage of the transition, where the frequency slowly decays to the Rossiter mode's frequency. Contrary to \cite{brs2008}, the transition of this case does not seem to trigger other Rossiter modes (figure \ref{fig:transition_M050_fft}). The origin of this different behaviour is believed to reside in the different base flows employed. Note that here we make use of a steady exact flow solution as base flow, whereas, in their study, they employed an averaged flow-field. The average flow is not itself a steady solution, so, the flow reacts to that artificial flow as if it were a perturbation to an exact solution, triggering other leading Rossiter modes as seen also in subsection \ref{subsec:mom_t_and_s}.

\subsubsection{Stability Evolution across Mach Number}

\begin{figure}
  	\centering
    \input{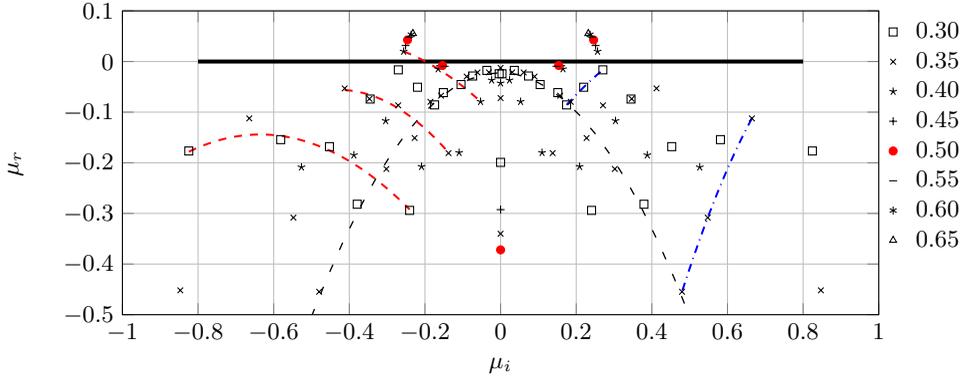}
    \caption{Eigenspectrum of the family of equilibrium solutions across Mach number. The red dashed lines show the rise of the unstable branch at Mach numbers 0.3, 0.35 and 0.4. The blue dash-dotted lines represent the evolution of a decaying branch as the Mach number is increased. The thin black dashed line highlights the branch appearing due to noise in the stable steady states.}
 	\label{fig:EGV_vs_M}
\end{figure}

In the present section, we evaluate how the changes in flow compressibility and the associated acoustic speed translate into the stability of these equilibrium solutions. In order to carry out this stability analysis across Mach number, we use the eigenspectrum and eigenmodes resulting from our adjoint Navier-Stokes framework and focus the present analysis on the receptivity.

\begin{figure}
 	\centering
    \input{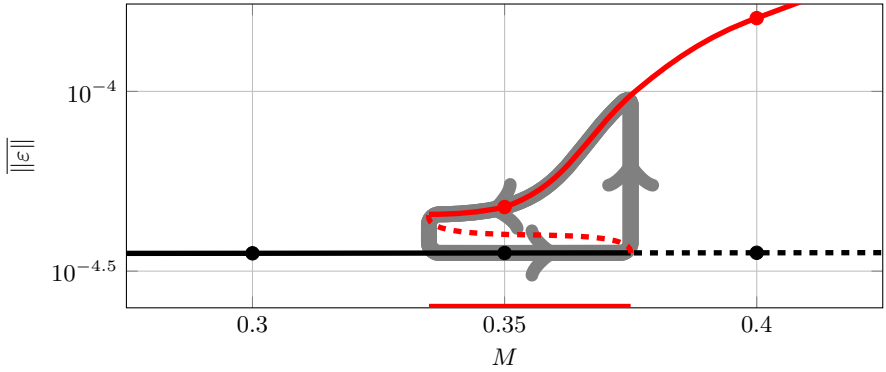}
 	\caption{Illustration of the subcritical Hopf bifurcation of the periodic and steady families of solutions across Mach number. The solid lines represent stable flow states, whereas the dashed lines show unstable flow configurations. The red and black coloured lines indicate the periodic and steady families of solutions, respectively. Also with the same colour coding, the circle symbols show the actual flow solutions computed in this investigation. The hysteresis loop is depicted with the thick grey solid lines, where the arrows show the loop direction. For this illustration, the Mach number range where bistability occurs is represented with the red coloured horizontal axis.}
	\label{fig:bifurcation}
\end{figure}

Figure \ref{fig:EGV_vs_M} shows the eigenvalue spectrum for the steady solutions from Mach numbers $0.30$ to $0.65$. For clarity, the eigenspectrum from the stable $M=0.25$ solution has not been included since all the eigenvalues corresponding to that Mach number lack dynamical importance ($\mu_r \lesssim -20 $). From this figure \ref{fig:EGV_vs_M}, notice that the steady solutions at $M=0.3$ and $M=0.35$ are stable, where all their eigenvalues are $\mu_r < 0$. Contrarily, all the other steady states are globally unstable since they have at least one eigenvalue with $\mu_r>0$. Similarly, \cite{brs2008} also observed that their 2D base flow was stable at $M=0.35$, but equivalently to the present investigation, it became unstable when raising the Mach number up to $M=0.6$, leaving the rest of parameters unchanged. Recalling the periodic orbits documented earlier, a stable periodic flow solution was found also at $M=0.35$ when descending in Mach number from $M=0.4$. This shows that both steady and periodic flow solutions are stable at this particular Mach number, where the flow will evolve to one or the other depending on the initial condition chosen. Additionally, this suggests that the transition from steady to periodic solutions does not occur at the same Mach number as the transition in the opposite direction. At the same time, this can be interpreted as the cavity flow being compressible enough to maintain the self-sustained oscillations (when descending in Mach number), but not enough to trigger them (when ascending).
The coexistence of these two stable solutions at the same Mach indicates bistability, which is characteristic of a subcritical Hopf bifurcation. Figure \ref{fig:bifurcation} shows an illustration of this type of bifurcation with the data points from the families of periodic and steady flow solutions. With ascending Mach number, the equilibrium solutions are stable up to the subcritical bifurcation point, located between Mach numbers 0.35 and 0.40. At this bifurcation, two unstable branches emerge, giving rise to unstable equilibrium and periodic solutions. Hence, far from following any of the unstable branches past the bifurcation point, the system undergoes a transition towards the corresponding stable limit cycle at the same Mach number. Once the flow reaches this periodic orbit, further variations in Mach number will displace the system along the family of periodic solutions as described earlier in section \ref{sec:periodic_family}. However, if the Mach number is reduced below the lower limit of the bistability range, the periodic orbit ceases to be stable, which leads to a rapid decay towards its corresponding steady state at the same Mach number. Thus, the bistability leads to a hysteresis loop around this bistable range (see figure \ref{fig:bifurcation}), which results into the two different transitions described above, from a steady state to a periodic orbit and vice versa. Unfortunately, in order to determine the exact Mach number at which these transitions occur, a finer sampling of flow solutions would be required.

Going back to figure \ref{fig:EGV_vs_M}, in an attempt to shed light on the origin of the instability as a function of Mach number, the leading eigenvalue can be traced back from the unstable to the stable regime, and there its eigenmode can be analysed. Note that the eigenvalues with a positive growth rate from the solutions ranging from $M=0.4$ to $M=0.65$ are remarkably aligned in the unstable plane. Intuitively, it might be tempting to relate these unstable eigenvalues to the leading (stable) eigenvalue from the $M=0.3$ solution, which is placed just below. If that were the case, the leading eigenvalue corresponding to the $M=0.35$ equilibrium solution should be placed somewhere in between; but this is not the case. Studying figure \ref{fig:EGV_vs_M} in more detail, we may observe the branch containing this leading eigenvalue at $M=0.3$ (highlighted by the blue dash-dotted line) has its analogous branch at $M=0.35$, which exhibits a further damped character, with also a much higher frequency. In addition, at this point, it is worth highlighting that the eigenvalues aligned in a parabolic shape at Mach numbers 0.3 and 0.35 are believed to be associated with the noise present in those simulations \citep{Bagheri:2014}.

\begin{figure}
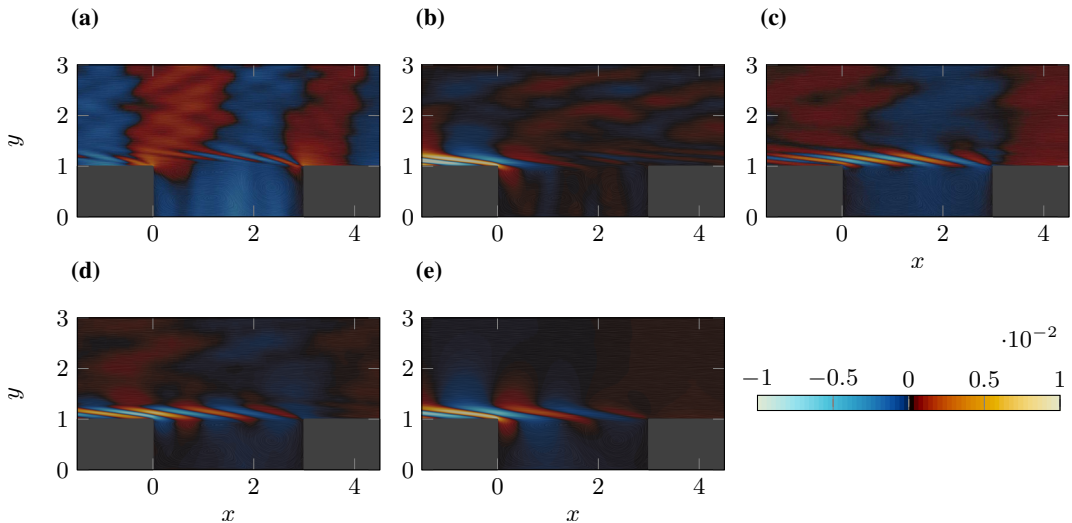

 \begin{subfigure}[b]{0.32\textwidth}
  	\centering
  	\subcaption{}
    \input{./M030_adj_umom_red_branch_1}
    \label{fig:M030_adj_lead1}
 \end{subfigure}
 ~
 \begin{subfigure}[b]{0.32\textwidth}
  	\centering
  	\subcaption{}
    \input{./M030_adj_umom_red_branch_2}
    \label{fig:M030_adj_lead4}
 \end{subfigure}
 ~
 \begin{subfigure}[b]{0.32\textwidth}
  	\centering
  	\subcaption{}
    \input{./M035_adj_umom_red_branch_1}
    \label{fig:M035_adj_lead1}
 \end{subfigure}\vspace{-0.25cm}\\
 \begin{subfigure}[b]{0.32\textwidth}
  	\centering
  	\subcaption{}
    \input{./M035_adj_umom_red_branch_2}
    \label{fig:M035_adj_lead2}
 \end{subfigure}
 ~
 \begin{subfigure}[b]{0.32\textwidth}
  	\centering
  	\subcaption{}
    \input{./M040_adj_umom_red_branch}
    \label{fig:M040_lead} 
 \end{subfigure}
 \begin{subfigure}{0.32\textwidth}
 	\vspace{-2.5cm}
 	\centering
  	\input{./unstable_egv_evolution_colourbar}
 \end{subfigure}
 \vspace{0.5cm}
 \caption{Evolution of the streamwise momentum receptivity of some of the eigenvalues from the unstable branch across Mach number. The first mode is shown in figures \textbf{(a)}, \textbf{(c)} and \textbf{(e)} at Mach numbers 0.30, 0.35 and 0.4, respectively. The fourth mode at $M=0.3$ and the second mode at $M=0.35$ are also shown in \textbf{(b)} and \textbf{(d)}.}
\label{fig:unstable_egv_evolution}
\end{figure}

Figure \ref{fig:EGV_vs_M} also shows a rising eigenvalue branch from $M=0.3$ to $M=0.4$, highlighted with red dashed lines. In contrast to the preceding blue dash-dotted decaying branch, this new branch exhibits a least stable nature as the Mach number of the steady solution is increased, moving towards a lower frequency regime. At $M=0.4$ its leading eigenvalue presents a positive growth rate, sitting in the unstable plane. For reference, this branch is also denoted as the unstable branch, where its eigenvalues are sorted numerically with decreasing characteristic frequency. Figure \ref{fig:unstable_egv_evolution} displays the evolution of some of the eigenvalues from this unstable branch across Mach number from a streamwise momentum receptivity point of view. Respectively, the first and fourth modes at $M=0.3$ are shown in figures \ref{fig:M030_adj_lead1} and \ref{fig:M030_adj_lead4}. There, the areas with high receptivity appear to displace continuously from the incoming laminar boundary layer towards the shear layer as the frequency increases along the branch. Additionally, as seen in figure \ref{fig:EGV_vs_M}, the eigenvalues from this branch get closer together as the Mach number is raised. As reflected in figures \ref{fig:M035_adj_lead1} and \ref{fig:M035_adj_lead2}, this phenomenon also occurs in their respective eigenvectors, where the first and second adjoint modes at $M=0.35$ show a very similar receptivity (both amplitude and shape-wise) in the streamwise momentum component. Hence, if this approaching behaviour would continue, the first and second modes would merge eventually. This would result into only three eigenvalues forming the unstable branch, which is precisely the case observed at $M=0.4$. Unfortunately, a finer sampling of equilibrium solutions across Mach number would be required in order to rigorously demonstrate this potential eigenvalue merging, which is believed to have a relevant influence on the onset of the unstable eigenvalue. 

Lastly, figure \ref{fig:M040_lead} reveals the adjoint streamwise momentum component of the unstable mode at $M=0.4$. Inspecting the progression of this particular adjoint mode across Mach number, we detect how the dominant activity of this mode moves gradually from the shear layer towards the proximity of the leading edge and incoming boundary layer, where it halts. This phenomenon agrees with the adjoint momentum components of the unstable mode at $M=0.5$ as shown earlier in figure \ref{fig:M050_modes_and_wavemaker}, which exhibit a remarkably similar receptivity to the corresponding mode at Mach number 0.4. Perhaps the only noticeable difference between figures \ref{fig:M050_b} and \ref{fig:M040_lead} resides in the content of the free-stream, where a slightly more distinct downstream reflected wave appears with increasing flow compressibility. Note that an analogous Mach number effect was shown in the Rossiter modes taking place in time-periodic flow solutions. In the same manner, once the unstable mode of the steady solution is active, a higher flow compressibility results in a stronger upstream-travelling acoustic wave reflected off the trailing edge.
This phenomenon reinforces the acoustic-convective feedback loop, yielding a larger growth rate $\mu_r$. In addition, the close relationship amongst the periodic orbits and the stability of the steady solutions is manifested in the evolution of the frequency of this unstable mode. Similarly to the flow periodic trajectories, the non-dimensional frequency $\mu_i$ of the unstable mode decreases as the Mach number raises (see table \ref{table:freq_vs_M}). The reason behind this behaviour is simply the reduction in the propagating speed of sound relative to the convective flow velocity, which increases the lag of the acoustic-convective feedback loop. Relating these results to previous studies, the decreasing trend in the unstable mode's frequency as a function of Mach number was also observed in \cite{yamouni2013}. Analogously to our findings, they observed the appearance of new unstable modes as the Mach number was raised (five extra modes at $M=0.5$ and seven at $M=0.9$). In their investigation, the constant presence of unstable modes from purely convective origin is prone to exert a frequency and growth rate modulation over the exclusively compressible unstable modes, and vice versa. In fact, the proportional increase in growth rate as a function of Mach number was reported as surprising, when comparing their results with literature. Hence, in the present work, we have illuminated the physical mechanisms causing the increase in growth rate and the appearance of additional unstable modes with Mach number, from an exclusively compressible receptivity point of view.

\begin{table}
\centering
\begin{tabular}{cccc}
 \hline
 \hspace{0.5cm}$M_\infty$\hspace{0.5cm} & $St_{periodic}$  & $\mu_i$ & $\mu_r$ \\ \hline \hline
 0.30 & $-$ & \hspace{0.5cm}0.824514\hspace{0.5cm} & -0.176693\\
 0.35 & \hspace{0.5cm}0.24871\hspace{0.5cm} & 0.412341 & -0.053106 \\
 0.40 & 0.24472 & 0.256365 &  \hspace{0.5cm}0.019984\hspace{0.5cm} \\
 0.45 & 0.23923 & 0.251265 &  0.031612 \\
 0.50 & 0.23355 & 0.246062 &  0.042644 \\
 0.55 & 0.22822 & 0.241150 &  0.048355 \\
 0.60 & 0.22319 & 0.236504 &  0.052891 \\
 0.65 & 0.21792 & 0.231924 &  0.055347 \\
 \hline
\end{tabular}
\caption{Comparison across Mach number of the non-dimensional frequency associated with the periodic solutions and the first eigenvalue of the unstable branch of the corresponding steady solution.}
\label{table:freq_vs_M}
\end{table}

\section{Conclusions}

A newly developed framework to compute steady and periodic compressible flow solutions has been successfully applied on a two-dimensional open cavity flow. With this method, we have computed a family of compressible periodic flow solutions across Mach number for the first time. We have also calculated a family of equilibrium solutions using the same open cavity configuration. The Reynolds number based on the cavity depth was carefully chosen to avoid any convective instabilities in the flow-field, restricting the system to be only driven by purely compressible self-sustained oscillations. With this setup, we were able to show how the two families of periodic and steady solutions collapse in the quasi-incompressible regime ($M\approx0.30$), which proves that the flow compressibility has a destabilising effect in cavity flows.

To shed light on this phenomenon a thorough analysis of the evolution of the compressible events across Mach number has been carried out. Particular emphasis was put on the flow-acoustic shear layer interaction, which dominates the dynamics. This interaction was shown to have a major effect on the overall sound radiation.
Furthermore, these periodic flow solutions are stable even under large disturbances, unlike the steady solutions, which transition towards the corresponding limit cycle.
To analyse the bifurcation point, a detailed stability analysis of the equilibrium solutions was also carried out. Dynamic mode decomposition of an adjoint simulation snapshot sequence was used to recover the adjoint global modes and eigenspectrum. The forward global modes were also similarly found for the steady solution at $M=0.5$, using DMD over a non-linear forward dataset, assuming linear flow behaviour in the initial stages of time marching. 
The forward approximated modes revealed the laminar shear layer as the leading amplifier of the instability. The respective adjoint modes highlighted the leading edge and incoming boundary layer as the regions with the highest receptivity.

Additionally, the evolution of the frequency corresponding to the unstable mode was tracked throughout the steady-to-periodic transition at $M=0.5$. This showed that the system quickly leaves the linear regime, even before the shear layer and downstream vortices begin the merging process. The wave maker region was computed showing activity in the boundary layer and shear layer, with maximum intensity at the leading edge, only to decay towards the trailing edge.

The evolution of the eigenvalue instability was analysed across Mach number. This analysis showed that the transition from steady to periodic flow and vice versa do not occur at the same Mach number, indicating a subcritical Hopf bifurcation of the steady and periodic families. This shows that locally stable periodic and steady solutions may coexist over a short Mach number range.
The eigenvalue branch which eventually becomes unstable as the Mach number is raised was also identified. The lack of both forward modes and a finer sampling of solutions across Mach number permits us only to hypothesise on the onset of the unstable character, and analyse such modes from a receptivity point of view. Hence, a forward stability analysis and a finer sampling of steady flow solutions (in particular from $M=0.35$ to $M=0.40$) are indicated for future work. A dataset containing both families of solutions is available online \citep{cavity_sol_dataset}.

\appendix
\section{}
\label{app:gov_eq}

The in-house code used to produce the DNS data presented above solves the full compressible Navier-Stokes equations. This set of equations are derived by applying the mass conservation, Newton's second law (momentum conservation) and the conservation of energy, resulting into a non-linear set of five partial differential equations (PDEs), known as continuity (\ref{eq_g:conteq}), momentum (\ref{eq_g:momeq}) and total energy (\ref{eq_g:enereq}) equations, which are defined for the entire flow domain $\Omega$. In particular, the equations are used in their conservative form, where the variables are density $\rho$, the three-dimensional momentum vector $\rho \vec{u}$ and the conserved energy $\rho E$ 
\begin{gather}
\label{eq_g:conteq}
\frac{\partial \rho}{\partial t} + \nabla \cdot \left(\rho\vec{u}\right)=0 \\
\label{eq_g:momeq}
\frac{\partial \rho \vec{u}}{\partial t} + \nabla \cdot \left(\rho \vec{u} \otimes \vec{u} \right) = -\nabla p +\nabla \cdot \tau +\rho \vec{f} \\
\label{eq_g:enereq}
\frac{\partial \rho E}{\partial t} + \nabla \cdot \left(\rho E \vec{u}\right) = -\nabla \cdot \left(p \vec{u}\right) + \nabla \cdot \left(\tau \cdot \vec{u}\right) + \nabla \cdot \vec{q} +\rho \vec{u} \cdot \vec{f},
\end{gather} 
where $\vec{u}$ is the velocity vector, $\vec{f}$ represents an external volume force, and $E$ is the total energy defined as the sum of internal and kinetic energy $E=e_{int}+\frac{1}{2}\vec{u}\cdot \vec{u}$. The equation of state (\ref{eq_g:eos}) gives closure to the system linking pressure $p$ with density $\rho$, temperature $T_{e}$ and internal energy $e_{int}$\begin{equation}
\label{eq_g:eos}
p = \frac{\rho T_{e}}{\gamma M_\infty^2} = (\gamma - 1)\rho e_{int}.
\end{equation} 

The shear stress tensor $\tau$ is symmetric and is defined as \begin{equation}
\tau = \frac{\mu}{Re_\infty}\left(\nabla\vec{u}+\left(\nabla\vec{u}\right)^{\text{T}}\right) - \left(\frac{2}{3}\frac{\mu}{Re_\infty}\nabla \cdot \vec{u}\right) \delta,
\end{equation} with $\mu$ as dynamic viscosity and $\delta$ as the identity matrix. The term $\vec{q}$ in (\ref{eq_g:enereq}) is known as heat flux vector, which is defined as \begin{equation}
\vec{q} = \frac{\mu}{Re_\infty \left(\gamma-1\right)M_\infty^2Pr_\infty}\nabla T_e
\end{equation} where $\gamma$ is the isentropic coefficient of the fluid which is assumed to be $1.4$. 

This set of equations are non-dimensionalised with the Reynolds $(Re)$, Mach $(M)$ and Prandtl $(Pr)$ numbers; which are defined as
\begin{equation}  Re = \frac{\rho_{\infty} u_{\infty} D}{\mu_{\infty}}, \ M = \frac{u_\infty}{\sqrt{\gamma R {T_{e}}_\infty}} \ \mathrm{and} \ Pr= \frac{\mu_\infty c_p}{\kappa_\infty}. 
\end{equation}
The sub-script $_\infty$ indicates free stream variables, and $D$ is the characteristic length of the problem (depth of the cavity), $R$ is the ideal gas constant $\left(287\left[m^2/Ks^2\right]\right)$, $c_p$ is the specific heat and $\kappa$ is the thermal conductivity. The Prandtl number is assumed to be constant at 0.72 in all the simulations presented in this report. The molecular viscosity $\mu$ is computed using Sutherland's law \citep{white} \begin{equation}\mu = T_e^{3/2}\left(\frac{1+R_{Su}}{T_e+R_{Su}}\right),
\end{equation} setting the ratio of the Sutherland constant over free-stream temperature to $0.36867$.

\section{}
\label{appA}

This appendix gathers the simulation parameters corresponding to the steady (table \ref{table:equilibria}) and periodic (table \ref{table:periodic}) exact flow solutions presented above. The parameters describing the boundary layer's momentum thickness in table \ref{table:periodic} are averaged values over the entire solution's period. Despite the fact that the computation of periodic flow solutions for $M\lesssim 0.30$ yielded steady states, we have run these two solutions further with the steady solution setup to ensure they satisfy the same convergence criterion than the other steady solutions. This is the reason of the slight differences in the simulation parameters presented tables \ref{table:equilibria} and \ref{table:periodic} at $M=0.25$ and $M=0.30$.

\begin{table}
\centering
\begin{tabular}{ccccccc}
 \hline
  \hspace{0.05cm}$M_\infty$\hspace{0.05cm} & $Re_D$  & $Re_\Theta$  & $L/\Theta$ && $\mu_r$ & $\mu_i$ \\ \hline \hline
 0.25 & 2000 & 50.51 & 118.77 && - & -\\
 0.30 & \hspace{0.05cm}2000\hspace{0.05cm} & \hspace{0.05cm}50.36\hspace{0.05cm} & \hspace{0.05cm}119.13\hspace{0.05cm} &\hspace{0.05cm}& \hspace{0.05cm}-0.176693\hspace{0.05cm} & \hspace{0.05cm}0.824514\hspace{0.05cm} \\
 0.35 & 2000 & 50.59 & 118.56 && -0.053106 & 0.412341 \\
 0.40 & 2000 & 49.00 & 122.52 && 0.019984 & 0.256365 \\
 0.45 & 2000 & 50.34 & 118.77 && 0.031612 & 0.251265 \\
 0.50 & 2000 & 50.52 & 118.74 && 0.042644 & 0.246062 \\
 0.55 & 2000 & 50.51 & 118.77 && 0.048355 & 0.241150 \\
 0.60 & 2000 & 47.79 & 125.52 && 0.052891 & 0.236504 \\
 0.65 & 2000 & 44.40 & 135.12 && 0.055347 & 0.231924 \\
 \hline
\end{tabular}
\caption{Parameters and stability of the 2D cavity equilibrium solutions. $D$ and $\Theta$ are the cavity depth and momentum thickness at the cavity's leading edge, respectively. The columns $\mu_r$ and $\mu_i$ contain the growth rate and frequency of the leading (first) eigenvalue of the unstable branch.}
\label{table:equilibria}
\end{table}

\begin{table}
\centering
\begin{tabular}{ccccc}
 \hline
  \hspace{0.25cm}$M_\infty$\hspace{0.25cm} & $Re_D$  & $Re_\Theta$  & $L/\Theta$ & $St$ \\ \hline \hline
 0.25 & 2000 & 50.50 & 118.82 & $-$ \\
 0.30 & \hspace{0.25cm}2000\hspace{0.25cm} & \hspace{0.25cm}50.36\hspace{0.25cm} & \hspace{0.25cm}119.14\hspace{0.25cm} & $-$  \\
 0.35 & 2000 & 50.55 & 118.68 & \hspace{0.25cm} 0.2489 \hspace{0.25cm} \\
 0.40 & 2000 & 49.10 & 122.19 & 0.2447 \\
 0.45 & 2000 & 50.38 & 119.08 & 0.2393 \\
 0.50 & 2000 & 50.36 & 119.15 & 0.2335 \\
 0.55 & 2000 & 49.75 & 120.58 & 0.2282 \\
 0.60 & 2000 & 46.34 & 129.46 & 0.2231 \\
 0.64 & 2000 & 42.84 & 140.05 & 0.2190 \\
 0.65 & 2000 & 41.83 & 143.40 & 0.2179 \\
 0.66 & 2000 & 41.48 & 144.64 & 0.2172 \\
 0.67 & 2000 & 40.76 & 147.20 & 0.2167 \\
 0.68 & 2000 & 40.23 & 149.15 & 0.2167 \\
 0.69 & 2000 & 39.32 & 152.32 & 0.2170 \\
 0.70 & 2000 & 38.47 & 155.96 & 0.2170 \\
 0.71 & 2000 & 37.54 & 159.80 & 0.2162 \\
 0.72 & 2000 & 36.31 & 165.25 & 0.2162 \\
 0.73 & 2000 & 34.54 & 173.69 & 0.2131 \\
 0.74 & 2000 & 32.73 & 183.34 & 0.2114 \\
 0.75 & 2000 & 30.88 & 194.28 & 0.2097 \\
 0.76 & 2000 & 29.32 & 204.60 & 0.2086 \\
 0.77 & 2000 & 28.31 & 211.91 & 0.2086 \\
 0.78 & 2000 & 27.10 & 221.41 & 0.2104 \\
 0.79 & 2000 & 24.96 & 240.42 & 0.2107 \\
 0.80 & 2000 & 23.06 & 260.24 & 0.2100 \\
 0.80-fd & 2000 & 64.37 & 93.20 & 0.2090 \\
 \hline
\end{tabular}
\caption{Simulation parameters of the 2D cavity periodic solutions. $D$ and $\Theta$ are the cavity depth and momentum thickness at the cavity's leading edge, respectively. $St$ shows the Strouhal number of each periodic trajectory.}
\label{table:periodic}
\end{table}


\bibliographystyle{jfm}
\bibliography{References.bib}

\end{document}